\documentclass[preprint]{emulateapj}

\usepackage{natbib}
\usepackage{amsmath}
\usepackage{graphicx}
\usepackage{wasysym}
\usepackage{txfonts}
\usepackage{xspace}
\usepackage{url}
\usepackage{textcomp}
\usepackage{multirow}
\usepackage{lipsum}

\newcommand{\her}{Her~X-1\xspace}

\newcommand{\swift}{\textsl{Swift}\xspace}

\newcommand{\inte}{\textsl{INTEGRAL}\xspace}
\newcommand{\chandra}{\textsl{Chandra}\xspace}
\newcommand{\xmm}{\textsl{XMM-Newton}\xspace}
\newcommand{\sax}{\textsl{BeppoSAX}\xspace}
\newcommand{\suz}{\textsl{Suzaku}\xspace}

\newcommand{\xte}{\textsl{RXTE}\xspace}

\newcommand{\nustar}{\textsl{NuSTAR}\xspace}
\newcommand{\highe}{\texttt{highecut}\xspace}
\newcommand{\fdcut}{\texttt{FDcut}\xspace}
\newcommand{\npex}{\texttt{NPEX}\xspace}
\newcommand{\gabs}{\texttt{gabs}\xspace}
\newcommand{\cyclabs}{\texttt{cyclabs}\xspace}

\newcommand{\snr}{S/N\xspace}

\newcommand{\msun}{\ensuremath{\text{M}_{\odot}}\xspace}

\newcommand{\redchi}{\ensuremath{\chi^{2}_\text{red}}\xspace}

\newcommand{\feka}{\ensuremath{\mathrm{Fe}~\mathrm{K}\alpha}\xspace}

\shorttitle{The cyclotron line  in Her~X-1}
\shortauthors{F\"urst et al.}
\slugcomment{Accepted in ApJ.}

\begin{document}


\title{The smooth cyclotron line  in HER~X-1 as seen with \textsl{NUSTAR}}


\author{Felix F\"urst\altaffilmark{1}}
\author{Brian W. Grefenstette\altaffilmark{1}}
\author{R\"udiger Staubert\altaffilmark{2}}
\author{John A. Tomsick\altaffilmark{3}}

\author{Matteo Bachetti\altaffilmark{4,5}}
\author{Didier Barret\altaffilmark{4,5}}
\author{Eric C. Bellm\altaffilmark{1}}
\author{Steven E. Boggs\altaffilmark{3}}
\author{Jerome Chenevez\altaffilmark{6}}
\author{Finn E. Christensen\altaffilmark{6}}
\author{William W. Craig\altaffilmark{3,7}}
\author{Charles J. Hailey\altaffilmark{8}}
\author{Fiona Harrison\altaffilmark{1}}
\author{Dmitry Klochkov\altaffilmark{2}}
\author{Kristin K. Madsen\altaffilmark{1}}
\author{Katja Pottschmidt\altaffilmark{9,10}}
\author{Daniel Stern\altaffilmark{11}}
\author{Dominic J. Walton\altaffilmark{1}}
\author{J\"orn Wilms\altaffilmark{12}}
\author{William Zhang\altaffilmark{10}}

\altaffiltext{1}{Cahill Center for Astronomy and Astrophysics, California Institute of Technology, Pasadena, CA 91125}
\altaffiltext{2}{Institut f\"ur Astronomie und Astrophysik, Universit\"at T\"ubingen (IAAT), T\"ubingen, Germany}
\altaffiltext{3}{Space Sciences Laboratory, University of California, Berkeley, CA 94720, USA}
\altaffiltext{4}{Universit\'e de Toulouse; UPS-OMP; IRAP; Toulouse, France}
\altaffiltext{5}{CNRS; Institut de Recherche en Astrophysique et Plan\'etologie; 9 Av. colonel Roche, BP 44346, 31028 Toulouse cedex 4, France}
\altaffiltext{6}{DTU Space, National Space Institute, Technical University of Denmark, Elektrovej 327, 2800 Lyngby, Denmark}
\altaffiltext{7}{Lawrence Livermore National Laboratory, Livermore, CA 94550, USA}
\altaffiltext{8}{Columbia Astrophysics Laboratory, Columbia University, New York, NY 10027, USA}
\altaffiltext{9}{Center for Space Science and Technology, University of Maryland
Baltimore County, Baltimore, MD 21250, USA}
\altaffiltext{10}{CRESST and NASA Goddard Space Flight Center, Astrophysics Science
Division, Code 661, Greenbelt, MD 20771, USA}
\altaffiltext{11}{Jet Propulsion Laboratory, California Institute of Technology, Pasadena, CA 91109, USA}
\altaffiltext{12}{Dr. Karl-Remeis-Sternwarte and ECAP, Sternwartstr. 7, 96
049 Bamberg, Germany}

\begin{abstract}
\her, one of the brightest and best studied  X-ray binaries, shows a cyclotron resonant scattering feature (CRSF) near 37\,keV. This makes it an ideal target for detailed study with the \textsl{Nuclear Spectroscopic Telescope Array} (\nustar), taking advantage of its excellent hard X-ray spectral resolution. We observed \her three times, coordinated with \suz, during one of the high flux intervals of its 35\,d super-orbital period. This paper  focuses on the shape and evolution of the hard X-ray spectrum. The broad-band spectra can be fitted with a powerlaw with a high-energy cutoff, an iron line, and a CRSF.
We find that the CRSF has a very smooth and symmetric shape, in all observations and at all pulse-phases. We compare the residuals of a line with a Gaussian optical depth profile to a Lorentzian optical depth profile and find no significant differences, strongly constraining the very smooth shape of the line. Even though the line energy changes dramatically with pulse phase, we find that its smooth shape does not.
Additionally, our data show that the continuum is only changing marginally between the three observations. These changes can be explained with varying amounts of Thomson scattering in the hot corona of the accretion disk. 
The average, luminosity-corrected CRSF energy is lower than in past observations and follows a secular decline. The excellent data quality of \nustar provides the best constraint on the CRSF energy to date.

\end{abstract}


\keywords{accretion -- stars: neutron -- pulsars: individual (Her~X-1) -- X-rays: binaries}



\section{Introduction}

One of the most useful features to study the magnetic field of neutron star binaries are Cyclotron Resonant Scattering Features (CRSFs, cyclotron lines for short). These are produced in the hot accretion column of the neutron star and appear as absorption-line-like features in the continuum spectrum. Continuum X-rays are scattered out of our line of sight when interacting with the electrons confined to quantized Landau-levels due to the strong magnetic field of the neutron star \citep[see, e.g.,][for a detailed description]{schoenherr07a}. The energies of CRSFs depend directly on the magnetic field strength in the line forming region, while their shape is a function of the magnetic field geometry. This makes their measurement the only available method to directly measure the magnetic field strength close to the surface of a neutron star.   

In the last few years it has become evident that the cyclotron line shape, depth, and, most of all, energy, can vary with pulse phase and overall luminosity \citep[for an overview see][]{caballero12a}. The pulse phase dependence results from viewing different parts of the accretion column and, assuming a dipole magnetic field, possibly the two different magnetic poles during one rotation. 
The luminosity dependence is thought to be related to different accretion rates, which move the line forming region in the accretion column to higher or lower altitudes above the neutron star surface, resulting in different magnetic field strengths. By measuring the changes of the CRSF, the physical conditions close to the neutron star can be inferred.

CRSFs were first discovered in \her \citep{truemper78a}, a pulsating X-ray binary system with a $2.3\pm0.3$\,\msun optical companion, at a distance of $6.6\pm0.4$\,kpc \citep{reynolds97a}. Due to the changing illumination of the optical companion by the neutron star, the surface has a variable temperature and can be hotter than for an isolated star, resulting in a early B- or late A-type classification. 
A CRSF is clearly seen in the spectrum around 37\,keV, but its energy is strongly variable with pulse phase. In \inte data, \citet{klochkov08a} found that it changes from around 30\,keV in the off-pulse to over 40\,keV during the main-peak of the pulse profile. It is now also firmly established that the line energy correlates positively with the X-ray luminosity \citep{staubert07a, vasco11a, staubert13b}. \her's high luminosity and very strong CRSF line make it an ideal target for a detailed study of the line's behavior.

Besides the rotational period of the neutron star of $P\approx 1.238$\,s, two other periods are very important in the \her system: the orbital period of $P_\text{orb} = 1.7$\,d and a 35\,d cycle of the warped and precessing accretion disk \citep{giacconi73a}. 
As the system is seen almost edge on \citep[$85^\circ$ inclination,][]{scott00a}, the neutron star is eclipsed by its companion for about 20\% of every orbit. Additionally the accretion disk obstructs our view of the compact object twice per 35\,d cycle, dividing the light-curve into a ``Main-On", showing the highest X-ray fluxes, and a ``Short-On" phase, lasting for about 10--11 and 5--7 days, respectively \citep{jones76a, staubert13a}. In time-resolved studies the spectral parameters seem to change over the 35\,d  orbit, most remarkably during the ``Main-On'' phase \citep[see, e.g.,][]{ji09a, zane04a, vasco13a}. For a comprehensive understanding of the system all three periods must be taken into account.
%

With a high sensitivity above 10\,keV and an energy resolution of 1\,keV at 68\,keV, the \textsl{Nuclear Spectroscopic Telescope Array} \citep[\nustar; ][]{harrison13a} has brought  the investigation of CRSFs to a new level. \nustar consists of two identical telescopes, focusing X-rays between 3.5--79\,keV using depth-graded multi-layer grazing incidence optics with a 10\,m focal length. At the focal plane of each telescope a CZT detector with $64\times64$ pixels is located, called Focal Plane Module A and B (FPMA and FPMB), respectively.  
Compared to previous missions, the focusing and imaging capabilities allow for a much lower and simultaneously measurable  background, decreasing systematic uncertainties at high energies.
With relative timing capabilities of 1\,ms, \nustar is well suited to perform phase-resolved spectroscopy of pulse periods on the order of one second \citep[see also][]{mori13a}.

Since the cyclotron line at 37\,keV is perfectly suited for study by \nustar, \her was selected as one of the first binary science targets.
 All observations were coordinated with \suz \citep{mitsuda07a}, providing simultaneous soft X-ray coverage.
In this paper we focus on the analysis of the hard X-ray spectrum and the behavior of the CRSF with pulse phase and 35\,d phase. A comprehensive analysis of the combined \suz and \nustar data will be presented in a forthcoming publication. 

The remainder of the paper is structured as follows: Section~\ref{sec:obs} gives an overview of the observations, and the data reduction and analysis methods. In Sect.~\ref{sec:lc} the light curves are presented and the energy-resolved pulse profile is discussed. Phase-averaged spectral analysis of the three observations is presented in Sect.~\ref{sec:phasavgspec}, while Sect.~\ref{sec:phasresspec} describes the phase-resolved spectroscopy. In Sect.~\ref{susec:peakspec} we focus on spectroscopy of the peak of the pulse profile and investigate the evolution with 35\,d phase. In Sect.~\ref{sec:outlook} the results are summarized and an outlook to future work is given. Throughout the paper uncertainties are given at the 90\% confidence level ($\Delta \chi^2=2.7$ for one parameter of interest), unless otherwise noted.

%

\section{Observations and data reduction}
\label{sec:obs}
\begin{figure}
\plotone{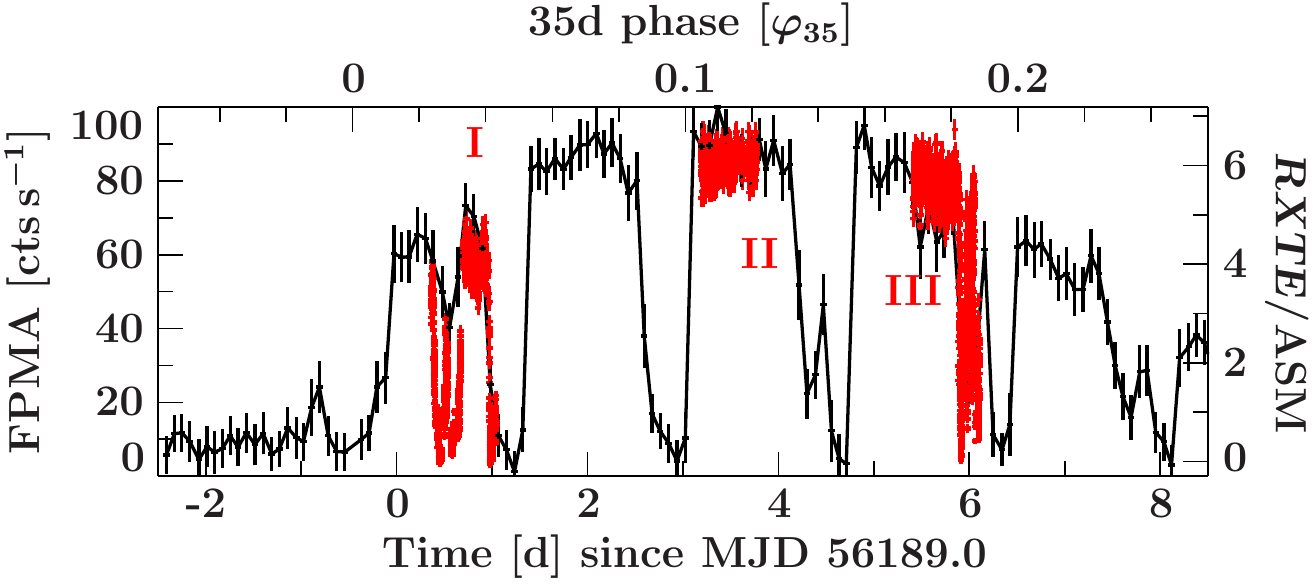}
\caption{
\xte/ASM data between 2--12\,keV in cts\,s$^{-1}$ folded onto the 35\,d period \citep{klochkov06a}, according to the top $x$-axis and right $y$-axis. \nustar FPMA data with 100\,s time resolution are over plotted in red, according to the left $y$-axis. The roman numerals indicate the three different observations, during the turn-on (I), main-on (II), and turn-off (III).}
\label{fig:batlc}
\end{figure}

To study the geometry of the accretion disk and the spectral changes over the ``Main-On'' phase in detail, three observations were obtained, each separated by roughly 2\,d to sample different phases of one ``Main-On'',
 as shown in Fig.~\ref{fig:batlc}. The first observation (I) was performed during the turn-on, the second during the brightest part (main-on, II), and the last one during the turn-off (III). 
Typically the turn-on of the main-on happens rather suddenly within one binary orbit, while the turn-off is a more gradual decline in flux over a few orbits \citep{scott00a}. This behavior is seen in the folded \xte/ASM lightcurve (Fig.~\ref{fig:batlc}) and also in the fluxes of the \nustar observations. The ASM 35\,d profile is based on \citet{klochkov06a}, using only turn-ons around binary phase $\phi=0.2$. The turn-on  analyzed here occurred at $\phi=0.24$ (MJD 56188.95).
      See Table~\ref{tab:obsdates} for an overview of the data used and resulting exposure times.

\begin{deluxetable*}{cccccccc}
\centering
\tablecolumns{8}
\tabletypesize{\scriptsize}
\tablecaption{\nustar and \suz observations and exposure times. \label{tab:obsdates}}
\tablehead{ \colhead{No.}  & \colhead{ObsID \nustar} & \colhead{ObsID \suz} &  \colhead{MJD range} & \colhead{$\varphi_{35} $} & \colhead{exp. \nustar} & \colhead{exp. XIS} & \colhead{exp. HXD} \\
&  \colhead{ (30002006XXX )} &  \colhead{ (4070510XX) }& & &\colhead{ FPMA / B [ks]} & \colhead{[ks]}  & \colhead{[ks]}}
\startdata 
I& 002 & 10  & 56189.362 --   56189.964 & 0.023--0.043 & 10.66 / 10.73 & 9.45  & 9.05 \\
II & 005 & 20 & 56192.194 -- 56192.773 &  0.105--0.121 & 21.87 / 22.13  & 22.61 & 18.72\\
III & 007 & 30 & 56194.395 -- 56194.889 & 0.169--0.182  & 17.77 / 17.88 & 15.91 & 14.75
\enddata
\tablecomments{The same time intervals where used for \nustar and \suz, but due to the different occultation times, the exposure times are not the same.}
\end{deluxetable*}

\subsection{\nustar}
The \nustar data were reduced using the standard NuSTARDAS pipeline v1.1.1. Spectra and light curves of Her~X-1 were extracted from a source region of 120" radius around the source location. Background spectra were extracted from a region with 80" radius as far away from the source as possible. Extracting the background in this way means that it was taken from another detector and in a region of different aperture flux background \citep[in prep.]{wik13a}. We therefore checked the validity of the background by scaling a nearby deep field observation to the observed 80--120\,keV background flux using the \texttt{nulyses} task \citep[in prep.]{zoglauer13a}. We found that the choice of the background did not significantly influence the spectral parameters, as \her is still a factor of 10 brighter than the background at the highest energies (and more than a factor of 100  brighter at low energies). As seen in Fig.~\ref{fig:batlc}, average count-rates were around 60, 90, and 80\,cts\,s$^{-1}$, for observation I, II, and III, respectively, while the background count-rate was around 0.06\,cts\,s$^{-1}$ in all observations.

Around 10\,keV, weak residuals from the tungsten L-edge of the optics of \nustar are present in the current version of the response. This feature depends on off-axis angle and is difficult to calibrate perfectly. As \her has a very high \snr, we see an absorption edge like feature in the phase-averaged spectra, see Fig.~\ref{fig:spec_highe_005}. To avoid the fit being influenced by this feature, we excised the 10--14.5\,keV energy range in the phase-averaged spectra. Aside from this feature, the \nustar responses are well calibrated and cross-checked by various simultaneous observations with other X-ray observatories \citep{harrison13a, madsen13a}.
 For all fits we used \nustar spectra between between 5--79\,keV, separately for FPMA and FPMB.

\subsection{\suz}
We reduced data from the \suz X-Ray Imaging Spectrometer \cite[XIS, ][]{koyama07a} using the standard pipeline as distributed with HEASOFT v6.13. Auxillary response files (ARFs) were created using \texttt{xissimarfgen} with a limiting number of 400,000 photons. We combined $2\times2$, $3\times3$, and $5\times5$ editing modes where available before extracting spectra and lightcurves.
We carefully checked for pile-up and removed the inner part of the PSF  following roughly the 3\% pile-up contours. We used two elliptical exclusion regions to do so, with radii $49''\times20''$ and $14''\times46''$ for observation I, $90''\times45''$ and $107''\times34''$ for observation II, and $78''\times32''$ and $31''\times73''$ for observation III in the east-west and north-south direction, respectively. As these regions exclude a large part of the chip, we decided to use a box shaped source region around the excluded core, with sides of $240''\times400''$.
However, this means that we extract data mainly from the outer parts of the chip, for which the contamination layer has higher uncertainties. The contamination layer can significantly influence the spectral shape, therefore it is important to know its thickness precisely. According to \citet{yamada12a} this layer is best understood for XIS\,3. We therefore used only this instrument in the current analysis, which focuses on higher energies and the cyclotron line.
 As the data were taken in the 1/4-window mode, no empty background region could be found on the chip. However, since the source is at least a factor of 20 brighter than the background at all times, background subtraction was not necessary.
We rebinned the spectra as described in \citet{nowak11a} and ignored the energy range around the know calibration features at 1.8\,keV and 2.2\,keV. The XIS data were fitted in the 0.8--8.5\,keV energy range.

We reduced data from the Hard X-ray Detector of \suz \citep[HXD, ][]{takahashi07a} with the standard pipeline using calibration files as published with HXD~CALDB\,20110913. Spectra were extracted using the tools \texttt{hxdpinxbpi} and \texttt{hxdgsoxbpi} for PIN and GSO, respectively. We obtained the tuned background model from the \suz website, as well as the recommended additional ARF for the GSO. PIN data were fitted between 20 and 70\,keV, GSO between 50 and 100\,keV. PIN data were rebinned to a \snr of 6 between 10-40\,keV and 3 above that. GSO data were not rebinned to follow the grouping of the background spectrum.
We added 1\% systematic uncertainties to the PIN data and 3\% to the GSO data to account for uncertainties in the background modeling.

\subsection{Cross-calibration of the instruments}

When comparing the \nustar and \suz/XIS fluxes, we find that the XIS values are up to 20\% below the \nustar values (see Tab.~\ref{tab:phasavg_005} in Sect.~\ref{sec:phasavgspec}). This is likely an effect of the large extraction region we used in the XIS data. This large region means we are using data only from the outer wings of the PSF, where the absolute flux calibration has higher uncertainties. To evaluate this, we extracted XIS spectra from all three XIS chips for observation II from a circular region with $120''$ radius and excised only the innermost core of the PSF using a circle with $30''$ radius. For XIS\,3, the flux difference then drops to only $\sim$10\%. However, the data showed systematically different slopes between the XIS detectors and compared to \nustar, likely an effect from remaining pile-up outside the small excised circle. The data from the large extraction region provided the best agreement  (other than the normalization) between all instruments.



Besides the uncertainty of the pile-up correction in XIS, the fact that the latest \nustar effective area was corrected using the Crab spectrum also contributes to differences in the flux measurement. For this calibration effort, a canonical spectrum of the Crab with $\Gamma = 2.1$ and a normalization of $A=10$\,ph\,cm$^{-2}$\,s$^{-1}$\,keV$^{-1}$ at 1\,keV was assumed \citep{madsen13a}. The spectrum measured by XIS\,3, however, is $\Gamma = 2.082\pm0.017$ and $A=9.33^{+0.24}_{-0.23}\,$ph\,cm$^{-2}$\,s$^{-1}$\,keV$^{-1}$ \citep{maeda08a}, i.e, a little lower than the assumed spectrum for \nustar. This difference can account for around a 10\% difference. A detailed description of the \nustar calibration will be given in \citet{madsen13a}.

The normalization difference between \nustar and PIN is on the order of 20\%. This is higher than expected as typically the cross-normalization between PIN and XIS should be about 15--20\%, while these values lead to an 40--50\% increased flux measurement of PIN compared to XIS. We attribute this discrepancy again to the higher uncertainties of the XIS extraction.
Forcing the normalization constant for PIN to be $1.19\times\textrm{C}_\text{XIS3}$ did not result in an acceptable fit ($\redchi > 3$).
 The GSO cross-normalizations are rather unconstrained due to the fact that \her is only barely detected in the instrument and the background is about a factor 5 higher than the source flux. The fluxes of the two independent \nustar detectors were within 4\% of each other in all cases.

Finally, a 2\,ks \swift/XRT snap-shot observation was done in parallel to observation I (see Fig.~\ref{fig:lc_all}). We extracted the data using the standard XRT pipeline and fitted the data between 0.8--9\,keV. The data are piled-up and due to the short observation do not contribute to constrain the model parameters. However, we checked the cross-calibration which comes out to $\text{C}_\text{XRT}=0.992\pm0.009$ compared to \nustar/FPMA. This value shows that \nustar provides fluxes in good agreement to other missions.








\section{Lightcurves and timing}
\label{sec:lc}
\begin{figure*}
\plotone{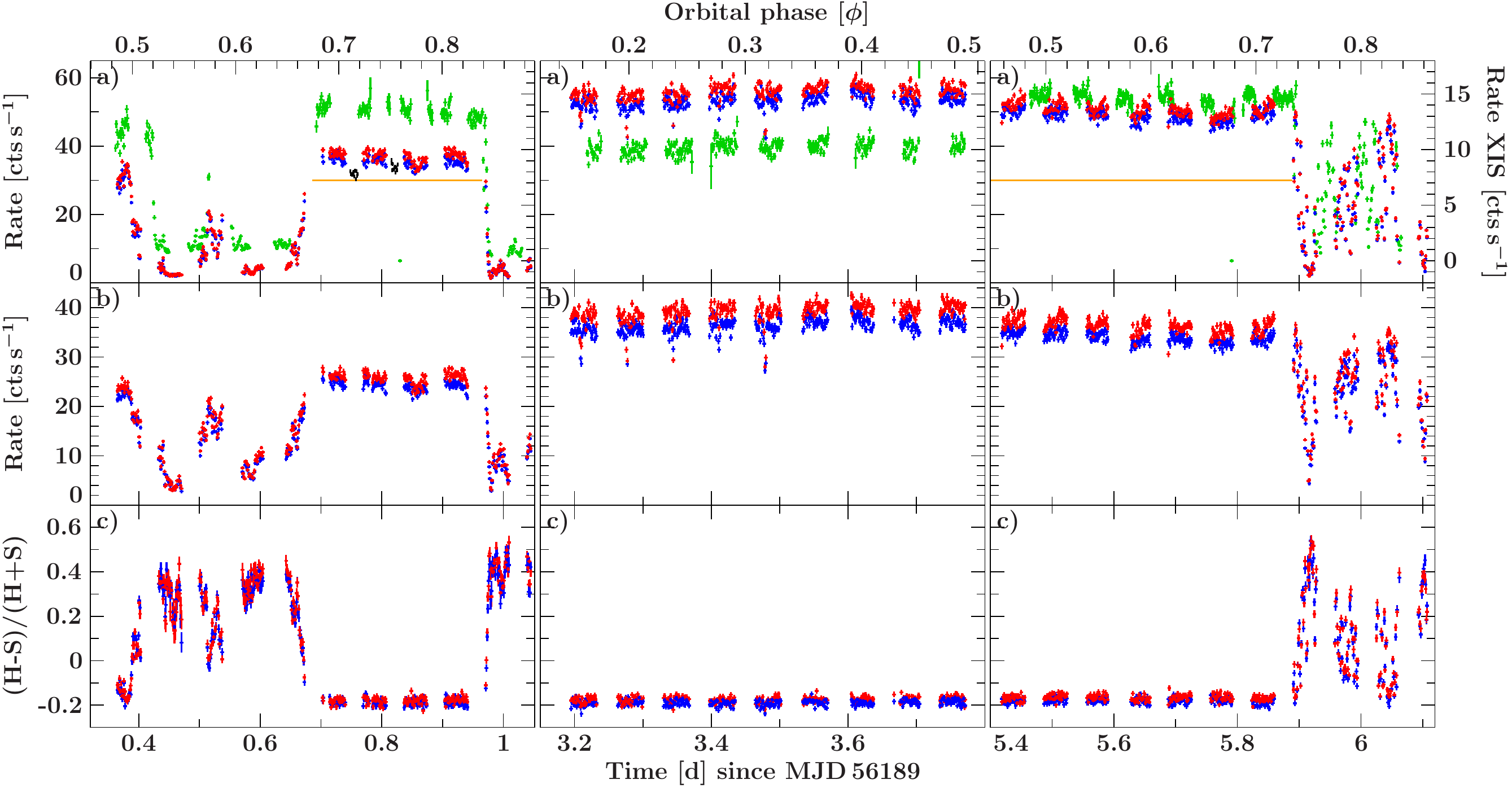}
\caption{\textsl{a)} Lightcurve of \nustar FPMA (red) and FPMB (blue) as well as XIS\,0 (green, right-hand $y$-axis) in the 3.5--10\,keV energy band with 100\,s resolution for observation I (left column), II (middle column), and III (right column). The orange horizontal lines indicates the times outside the intensity dips used to select the data for spectral analysis for observations I and III (see text for details). The black data points during observation I show the \swift/XRT data scaled according to the left-hand $y$-axis. \textsl{b)} Light curve of \nustar in the 10--79\,keV energy band. \textsl{c)} Hardness ratio between the light curves in the 3.5--10\,keV ($S$) and 10--79\,keV ($H$) energy bands. }
\label{fig:lc_all}
\end{figure*}

\her is known to show short intensity dips in the lightcurve, associated with increased absorption. These dips are known as anomalous dips and pre-eclipse dips \citep[see][and references therein]{igna11a} and are clearly seen in the \nustar lightcurves of observation I and III (Fig.~\ref{fig:lc_all}). Both energy bands, the soft 3.5--10\,keV and the hard 10--79\,keV band, show these dips, but with different intensity, so that in the hardness-ratio a strong hardening during the dips is observed. 
For the analyses in this paper, we disregarded all dip phases by filtering on the hardness ratio to use only data where the absorption was minimal, as indicated by the horizontal lines in Fig.~\ref{fig:lc_all}. The resulting reduced exposure times of the spectra are reflected in Tab.~\ref{tab:obsdates}. A detailed investigation of the dip spectra is beyond the scope of this paper.

Both satellites, \nustar and \suz, are in low-earth orbit and therefore the observed X-ray source is regularly occulted, resulting in the gaps in the light curves. As the orbits are not aligned, occultations happen not at the same time, so that strict simultaneity is not always achieved as, e.g., during observation I. As the orbits precess however, almost perfectly simultaneous observations where obtained during observation II. When describing the \nustar and \suz data in a simultaneous fit, short term spectral changes could result in different spectra for the both data-sets; however, the lightcurves do not indicate any such changes outside the intensity dips. The XIS count-rates change with respect to \nustar between observations due to the different exclusion regions used in the XIS reduction, i.e., for the brightest observation (II) the largest region was excised.

To obtain the pulse period, we barycentered the event times with the FTOOL \texttt{barycorr} using the DE-200 solar system ephemeris. We then  corrected the event times for the orbital motion of the neutron star, using the ephemeris by \citet{staubert09a}. We folded the combined event lists of all observations using the epoch folding method \citep{leahy87a} and  performed phase connection to estimate the uncertainty on the period. 
We found a pulse period of
\[P = 1.2377184353\pm0.000000020\,\mathrm{s}.\]
The period is consistent with results from the \textsl{Fermi}/GBM Pulsar Project \citep{finger09a}.

%
%

\begin{figure}
\plotone{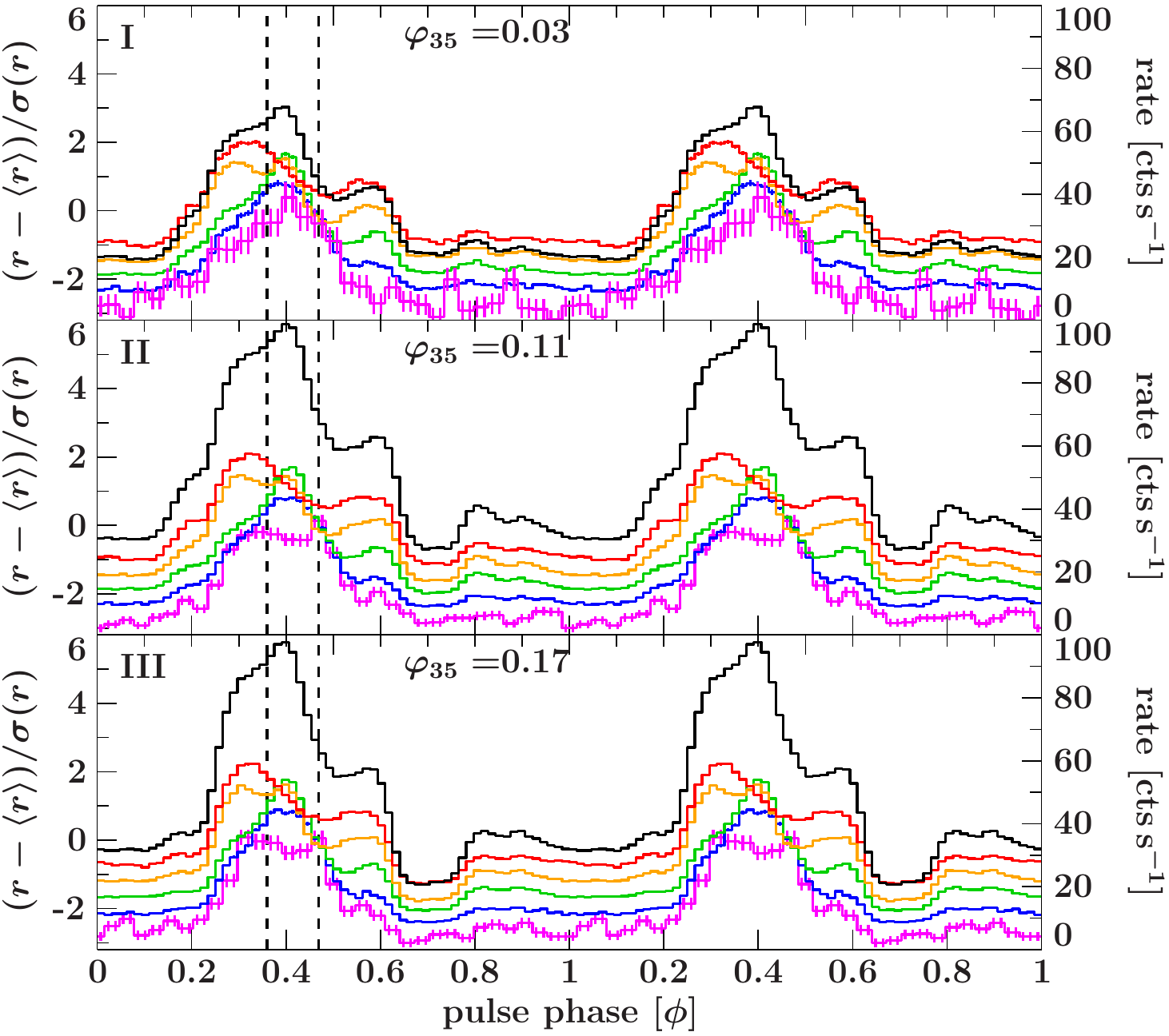}
\caption{Pulse profiles for observation I (top panel), II (center panel), and III (bottom panel). In black the average 3--79\,keV profile is shown, according to the right $y$-axis. Energy resolved profiles follow the left $y$-axis and are normalized by subtracting their average $\left<r\right>$ and dividing by their standard deviation $\sigma(r)$. Each consecutive profile is shifted by $-0.5$ in $y$ to increase visibility. The red profiles cover the 3--6\,keV energy range,  orange 6--10\,keV, green 10--20\,keV, blue 20--40\,keV, and magenta 40--79\,keV. The vertical dashed lines indicate the phase range used for the pulse-peak spectral analysis. The profiles are shown twice for clarity.}
\label{fig:pperg_all}
\end{figure}

The pulse profile shows a clear energy dependence, as well as a weak dependence on 35\,d phase, as shown in Fig.~\ref{fig:pperg_all} using 64 phase-bins. The main peak around $\phi = 0.4$ has  visible substructure at low energies, with the leading and trailing shoulder clearly separated. With increasing energy, both shoulders become less prominent, and instead the central core of the main peak more important. The inter-pulse peak around $\phi=0.8$ is very weak at this 35\,d-phase.  This is a well-known behavior in \her, see, e.g., \citet{scott00a}, and has been extensively studied in the literature \citep[see, e.g.,][]{deeter98a, kuster05a}. 
These energy-resolved pulse profiles are the basis for the phase-resolved spectroscopy described in Sect.~\ref{sec:phasresspec}.

\section{Phase-averaged spectroscopy}
\label{sec:phasavgspec}
\subsection {Continuum}
\label{susec:cont}
The continua of accreting neutron star binaries are typically modeled with phenomenological powerlaw models, with an high-energy cutoff.
As of now, there is no widely accepted and used model available which directly models the physical processes in the accretion column leading to the production of the hard X-ray continuum. Models are in development based on the work by \citet{becker07a} and have been used in special cases \citep[e.g., for 4U~0115+63,][]{ferrigno09a}, but still can not be generally applied. We therefore decided to describe the data with three of the most commonly used phenomenological models applied for \her:  high-energy cutoff (\highe), negative-positive exponential cutoff \citep[\npex,][]{makishima99a}, and Fermi-Dirac cutoff \citep[\fdcut,][]{tanaka86a}. 

All continuum models are based on a powerlaw with photon index $\Gamma$ and a high-energy cutoff with a folding energy $E_\text{fold}$. The \highe and \fdcut model also have the cutoff energy $E_\text{cut}$, at which the high-energy cutoff becomes relevant as a free parameter, while in the \npex model the third free parameter is the relative normalization $B_\text{cont}$ of the additional powerlaw component  with its photon  index frozen to $-2$. See also \citet{mueller13a} for a summary of their properties. Other similar phenomenological models like the \texttt{cutoffpl} or \texttt{comptt} did not lead to acceptable fits ($\redchi \gg 2$).

The \highe model is known to show a sharp feature at the cutoff energy, as discussed extensively in the literature
 \citep[see, e.g.,][]{kreykenbohm99a, kretschmar97b}. To smooth the transition between the pure powerlaw components and the cutoff-powerlaw, we multiplied the model with a line with a Gaussian optical-depth profile, with its energy tied to the cutoff energy (leaving as free parameters $\sigma_\text{c}$ and $\tau_\text{c}$), similar to the one described in \citet{coburn02a}. This model removes the feature in the residuals almost completely. However, due to \nustar's high sensitivity and spectral resolution, small features remain, though they do not influence the other model parameters and only result in a slightly worse fit in terms of \redchi.

For the broad-band \sax spectrum, \citet{dalfiume98a} found the best description to be a \highe continuum.
\citet{gruber01a} applied the \highe and \fdcut model to \xte observations of \her, and also find that the spectra are best described with the \highe model. The parameters of the cyclotron line, however, depended only weakly on the choice of the continuum. This very weak dependence is confirmed by \citet{klochkov08a} using \inte data. The modified \highe model is used in most of the recent work  on \her, see, e.g., \citet{staubert07a, klochkov08a, vasco13a}.

\begin{figure}
\plotone{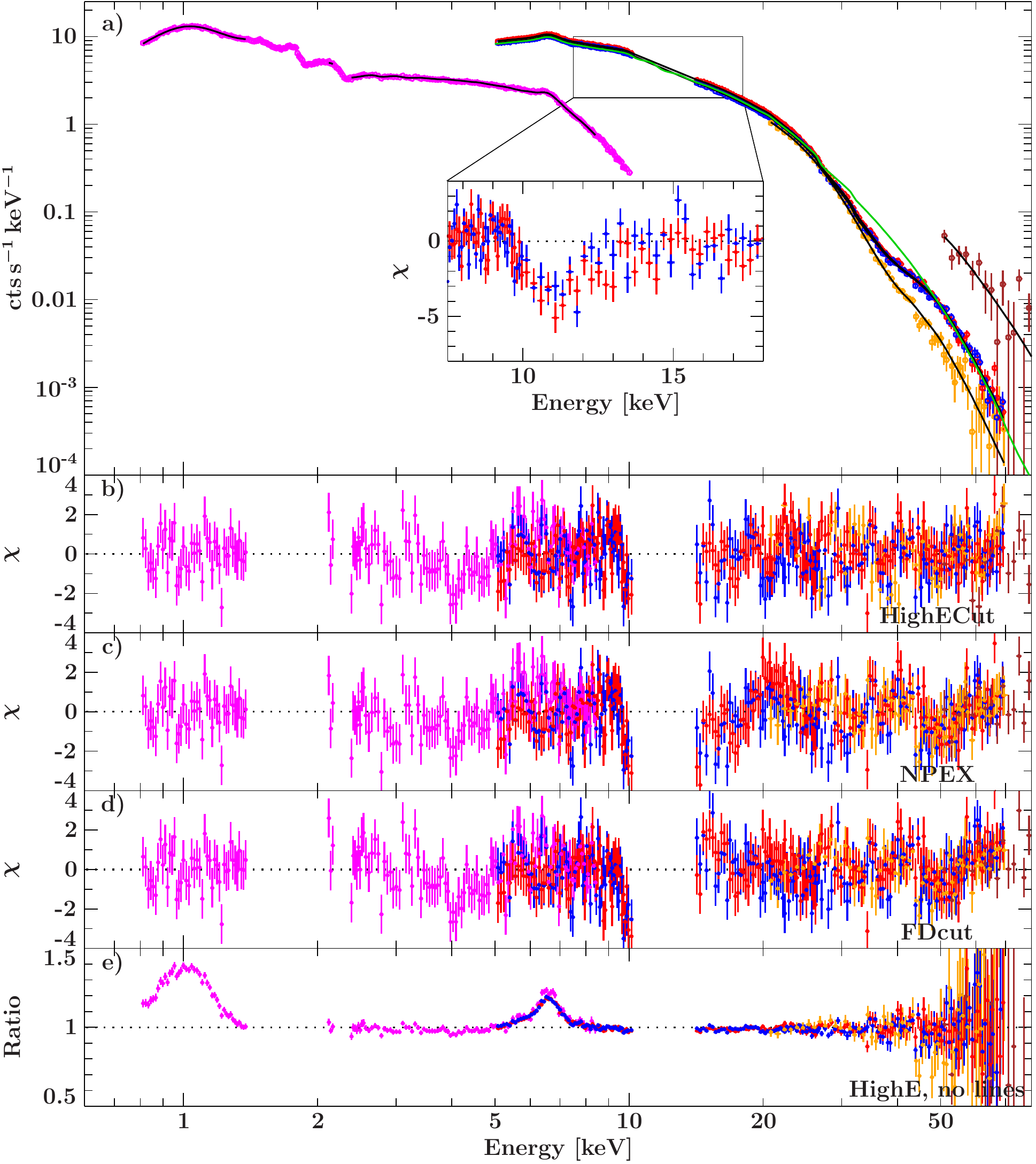}
\caption{\textsl{a)} \suz/XIS3 (magenta), FPMA (red), FPMB (blue), \suz/PIN (orange) and \suz/GSO (brown) spectra of observation II with the best-fit \highe model. The green line shows the model without the cyclotron line for FPMA. The inset shows the sharp residuals around the 10\,keV tungsten edge, disregarded in the fit. \textsl{b)} residuals in units of $\sigma$ for the best-fit \highe model; \textsl{c)} residuals for the \npex model; \textsl{d)} residuals for the \fdcut model; \textsl{e)} ratio of data over model for the \highe model without the fluorescence lines at 1\,keV and 6.4\,keV.}
\label{fig:spec_highe_005}
\end{figure}

We applied all three models to the three \nustar and \suz data-sets, the results are shown in Tables~\ref{tab:phasavg_002}, \ref{tab:phasavg_005}, and \ref{tab:phasavg_007}.
The \highe and \npex models result in a similar quality of fit for all three observations, while the \fdcut is typically a bit worse in terms of $\chi^2$ and clearly fails for observation III. The residuals from \npex, however, usually show a wavy structure in the hard \nustar band around the cyclotron line (see Fig.~\ref{fig:spec_highe_005}). We thus confirm earlier results that the preferred phenomenological model  to describe the broad \her spectrum is the \highe model, with the inclusion of an additional line to smear out the hard break at the cutoff energy.

\begin{deluxetable}{r|lll}
\tablecolumns{4}
\tabletypesize{\scriptsize}
\tablecaption{Fit parameters for observation I using three different continuum models.\label{tab:phasavg_002}}
\tablehead{ \colhead{Parameter}  & \colhead{HighE} & \colhead{NPEX} &  \colhead{FDcut}   }
\startdata 
 $ \Gamma$ & $0.913\pm0.007$ & $0.554^{+0.028}_{-0.023}$ & $0.888^{+0.008}_{-0.009}$ \\
 $ E_\text{cut} [\text{keV}]$ & $20.8^{+0.5}_{-0.6}$ & $5.83\pm0.09$ & $35.9^{+3.0}_{-2.3}$ \\
 $ E_\text{fold} [\text{keV}]$ & $10.15^{+0.28}_{-0.26}$ & -- & $6.0^{+0.5}_{-0.8}$ \\
 $ B_\text{cont}$ & -- & $\left(3.58\pm0.14\right)\times10^{-3}$ & -- \\
 $ E_\text{CRSF} [\text{keV}]$ & $38.3\pm0.6$ & $38.5\pm0.5$ & $38.4^{+0.7}_{-0.5}$ \\
 $ \sigma_\text{CRSF} [\text{keV}]$ & $6.5^{+0.5}_{-0.6}$ & $6.9\pm0.4$ & $8.4\pm0.4$ \\
 $ \tau_\text{CSRF}$ & $0.65\pm0.06$ & $0.91\pm0.06$ & $1.53^{+0.27}_{-0.19}$ \\
 $ \sigma_\text{c} [\text{keV}]$ & $2.54^{+0.24}_{-0.28}$ & -- & -- \\
 $ \tau_\text{c}$ & $0.156^{+0.016}_{-0.018}$ & -- & -- \\
 $ A_\text{BB}^\ddagger$ & $9^{+6}_{-5}$ & $30^{+8}_{-12}$ & $4.5^{+2.1}_{-4.6}$ \\
 $ kT [\text{keV}]$ & $0.111^{+0.017}_{-0.010}$ & $0.135^{+0.012}_{-0.019}$ & -- \\
 $ A(\text{Fe\,K}\alpha_\text{n})^\dagger$ & $0.28\pm0.05$ & $1.25^{+0.19}_{-0.18}$ & $0.28\pm0.05$ \\
 $ E(\text{Fe\,K}\alpha_\text{n}) [\text{keV}]$ & $6.504\pm0.020$ & $6.503\pm0.020$ & $6.508\pm0.022$ \\
 $ \sigma(\text{Fe\,K}\alpha_\text{n})[\text{keV}]$ & $0.155\pm0.029$ & $0.160^{+0.029}_{-0.028}$ & $0.152\pm0.029$ \\
 $ A(\text{Fe\,K}\alpha_\text{b})^\dagger$ & $0.75\pm0.09$ & $4.6^{+1.0}_{-0.7}$ & $0.66\pm0.08$ \\
 $ E(\text{Fe\,K}\alpha_\text{b}) [\text{keV}]$ & $6.43\pm0.08$ & $6.37^{+0.08}_{-0.09}$ & $6.39^{+0.07}_{-0.08}$ \\
 $ \sigma(\text{Fe\,K}\alpha_\text{b})[\text{keV}]$ & $0.90^{+0.13}_{-0.10}$ & $1.08^{+0.20}_{-0.15}$ & $0.80^{+0.10}_{-0.09}$ \\
 $ A(\text{Fe\,L}\alpha)^\dagger$ & $5.6^{+3.7}_{-2.1}$ & $18^{+17}_{-6}$ & $6.0^{+9.1}_{-2.6}$ \\
 $ E(\text{Fe\,L}\alpha) [\text{keV}]$ & $0.96^{+0.04}_{-0.06}$ & $0.97^{+0.04}_{-0.06}$ & $0.93^{+0.05}_{-0.10}$ \\
 $ \sigma(\text{Fe\,L}\alpha)[\text{keV}]$ & $0.143^{+0.030}_{-0.026}$ & $0.135^{+0.038}_{-0.024}$ & $0.149^{+0.052}_{-0.030}$ \\
 $ \mathcal{F}_\text{5--60\,keV}^\star$ & $3.193\pm0.010$ & $3.200\pm0.010$ & $3.192\pm0.010$ \\
 $ \text{C}_\text{FPMB}$ & $1.038\pm0.004$ & $1.038\pm0.004$ & $1.038\pm0.004$ \\
 $ \text{C}_\text{XIS\,3}$ & $0.878\pm0.005$ & $0.878\pm0.005$ & $0.880\pm0.005$ \\
 $ \text{C}_\text{PIN}$ & $1.207\pm0.014$ & $1.211\pm0.014$ & $1.205\pm0.014$ \\
 $ \text{C}_\text{GSO}$ & $1.9\pm0.4$ & $1.9\pm0.4$ & $1.9\pm0.4$ \\
$\chi^2/\text{d.o.f.}$   & 796.67/692& 821.19/694& 829.33/695\\$\chi^2_\text{red}$   & 1.151& 1.183& 1.193
\enddata
\tablecomments{$^\dagger$in $10^{-2}$ph\,s$^{-1}$\,cm$^{-2}$~;~$^\ddagger$in $10^{36}$\,erg\,s$^{-1}$ at 10\,kpc~;~$^\star$in keV\,s$^{-1}$\,cm$^{-2}$ } 
\end{deluxetable}

\begin{deluxetable}{r|lll}
\tablecolumns{4}
\tabletypesize{\scriptsize}
\tablecaption{Fit parameters for observation II using three different continuum models.\label{tab:phasavg_005}}
\tablehead{ \colhead{Parameter}  & \colhead{HighE} & \colhead{NPEX} &  \colhead{FDcut}   }
\startdata 
 $ \Gamma$ & $0.920\pm0.004$ & $0.584\pm0.012$ & $0.897\pm0.005$ \\
 $ E_\text{cut} [\text{keV}]$ & $20.68^{+0.27}_{-0.23}$ & $5.74\pm0.04$ & $35.6^{+1.3}_{-1.0}$ \\
 $ E_\text{fold} [\text{keV}]$ & $9.95\pm0.13$ & -- & $5.87^{+0.26}_{-0.30}$ \\
 $ B_\text{cont}$ & -- & $\left(3.60\pm0.07\right)\times10^{-3}$ & -- \\
 $ E_\text{CRSF} [\text{keV}]$ & $37.40^{+0.25}_{-0.24}$ & $37.73\pm0.20$ & $37.78^{+0.26}_{-0.24}$ \\
 $ \sigma_\text{CRSF} [\text{keV}]$ & $5.76^{+0.29}_{-0.27}$ & $6.36\pm0.18$ & $8.01\pm0.18$ \\
 $ \tau_\text{CSRF}$ & $0.614^{+0.028}_{-0.025}$ & $0.877^{+0.026}_{-0.025}$ & $1.49^{+0.12}_{-0.10}$ \\
 $ \sigma_\text{c} [\text{keV}]$ & $2.45^{+0.16}_{-0.15}$ & -- & -- \\
 $ \tau_\text{c}$ & $0.149\pm0.010$ & -- & -- \\
 $ A_\text{BB}^\ddagger$ & $6.8^{+1.6}_{-2.3}$ & $28^{+6}_{-9}$ & $7.5^{+1.4}_{-2.4}$ \\
 $ kT [\text{keV}]$ & $0.134^{+0.009}_{-0.019}$ & $0.140^{+0.009}_{-0.013}$ & -- \\
 $ A(\text{Fe\,K}\alpha_\text{n})^\dagger$ & $0.40\pm0.08$ & $1.68^{+0.15}_{-0.14}$ & $0.38^{+0.09}_{-0.08}$ \\
 $ E(\text{Fe\,K}\alpha_\text{n}) [\text{keV}]$ & $6.601^{+0.017}_{-0.016}$ & $6.601^{+0.014}_{-0.015}$ & $6.614^{+0.019}_{-0.018}$ \\
 $ \sigma(\text{Fe\,K}\alpha_\text{n})[\text{keV}]$ & $0.25\pm0.04$ & $0.253\pm0.020$ & $0.23\pm0.04$ \\
 $ A(\text{Fe\,K}\alpha_\text{b})^\dagger$ & $0.63\pm0.07$ & $3.94^{+0.26}_{-0.27}$ & $0.66^{+0.08}_{-0.09}$ \\
 $ E(\text{Fe\,K}\alpha_\text{b}) [\text{keV}]$ & $6.55\pm0.05$ & $6.48\pm0.04$ & $6.51^{+0.04}_{-0.05}$ \\
 $ \sigma(\text{Fe\,K}\alpha_\text{b})[\text{keV}]$ & $0.82^{+0.13}_{-0.10}$ & -- & $0.67^{+0.08}_{-0.07}$ \\
 $ A(\text{Fe\,L}\alpha)^\dagger$ & $6.0^{+4.1}_{-1.7}$ & $23^{+16}_{-7}$ & $6.3^{+3.6}_{-1.7}$ \\
 $ E(\text{Fe\,L}\alpha) [\text{keV}]$ & $0.957^{+0.028}_{-0.048}$ & $0.959^{+0.028}_{-0.047}$ & $0.961^{+0.027}_{-0.045}$ \\
 $ \sigma(\text{Fe\,L}\alpha)[\text{keV}]$ & $0.149^{+0.031}_{-0.021}$ & $0.149^{+0.031}_{-0.021}$ & $0.147^{+0.028}_{-0.019}$ \\
 $ \mathcal{F}_\text{5--60\,keV}^\star$ & $4.996\pm0.010$ & $5.011\pm0.010$ & $4.996\pm0.010$ \\
 $ \text{C}_\text{FPMB}$ & $1.0368\pm0.0020$ & $1.0368\pm0.0020$ & $1.0367\pm0.0020$ \\
 $ \text{C}_\text{XIS\,3}$ & $0.8219^{+0.0029}_{-0.0027}$ & $0.8198\pm0.0029$ & $0.8242^{+0.0029}_{-0.0028}$ \\
 $ \text{C}_\text{PIN}$ & $1.185\pm0.009$ & $1.188\pm0.009$ & $1.183\pm0.009$ \\
 $ \text{C}_\text{GSO}$ & $1.27\pm0.20$ & $1.25\pm0.20$ & $1.29\pm0.20$ \\
$\chi^2/\text{d.o.f.}$   & 820.43/711& 877.37/714& 934.67/714\\$\chi^2_\text{red}$   & 1.154& 1.229& 1.309
\enddata
\tablecomments{$^\dagger$in $10^{-2}$ph\,s$^{-1}$\,cm$^{-2}$~;~$^\ddagger$in $10^{36}$\,erg\,s$^{-1}$ at 10\,kpc~;~$^\star$in keV\,s$^{-1}$\,cm$^{-2}$ } 
\end{deluxetable}

\begin{deluxetable}{r|lll}
\tablecolumns{4}
\tabletypesize{\scriptsize}
\tablecaption{Fit parameters for observation III using three different continuum models.\label{tab:phasavg_007}}
\tablehead{ \colhead{Parameter}  & \colhead{HighE} & \colhead{NPEX} &  \colhead{FDcut}   }
\startdata 
 $ \Gamma$ & $0.916^{+0.004}_{-0.005}$ & $0.595\pm0.013$ & $0.889\pm0.005$ \\
 $ E_\text{cut} [\text{keV}]$ & $20.51^{+0.22}_{-0.20}$ & $5.84^{+0.06}_{-0.05}$ & $37.5^{+1.4}_{-1.2}$ \\
 $ E_\text{fold} [\text{keV}]$ & $10.19^{+0.17}_{-0.16}$ & -- & $5.40^{+0.28}_{-0.32}$ \\
 $ B_\text{cont}$ & -- & $\left(3.50\pm0.08\right)\times10^{-3}$ & -- \\
 $ E_\text{CRSF} [\text{keV}]$ & $38.5\pm0.4$ & $38.72^{+0.29}_{-0.28}$ & $38.3\pm0.4$ \\
 $ \sigma_\text{CRSF} [\text{keV}]$ & $6.4\pm0.4$ & $7.16^{+0.25}_{-0.24}$ & $8.24\pm0.17$ \\
 $ \tau_\text{CSRF}$ & $0.63\pm0.04$ & $0.95\pm0.04$ & $1.68^{+0.13}_{-0.12}$ \\
 $ \sigma_\text{c} [\text{keV}]$ & $2.00\pm0.16$ & -- & -- \\
 $ \tau_\text{c}$ & $0.123\pm0.009$ & -- & -- \\
 $ A_\text{BB}^\ddagger$ & $6.0^{+1.6}_{-2.5}$ & $25^{+7}_{-10}$ & $7.2^{+1.3}_{-2.0}$ \\
 $ kT [\text{keV}]$ & $0.144^{+0.012}_{-0.008}$ & $0.145^{+0.010}_{-0.008}$ & -- \\
 $ A(\text{Fe\,K}\alpha_\text{n})^\dagger$ & $0.54\pm0.05$ & $1.78^{+0.17}_{-0.16}$ & $0.64\pm0.06$ \\
 $ E(\text{Fe\,K}\alpha_\text{n}) [\text{keV}]$ & $6.582\pm0.015$ & $6.588\pm0.015$ & $6.586\pm0.014$ \\
 $ \sigma(\text{Fe\,K}\alpha_\text{n})[\text{keV}]$ & $0.272\pm0.020$ & $0.242\pm0.020$ & $0.295^{+0.022}_{-0.020}$ \\
 $ A(\text{Fe\,K}\alpha_\text{b})^\dagger$ & $0.50\pm0.08$ & $3.58^{+0.28}_{-0.29}$ & $0.33\pm0.09$ \\
 $ E(\text{Fe\,K}\alpha_\text{b}) [\text{keV}]$ & $6.47\pm0.08$ & $6.42\pm0.05$ & $6.27^{+0.12}_{-0.15}$ \\
 $ \sigma(\text{Fe\,K}\alpha_\text{b})[\text{keV}]$ & -- & -- & -- \\
 $ A(\text{Fe\,L}\alpha)^\dagger$ & $5.3^{+3.2}_{-1.5}$ & $22^{+13}_{-6}$ & $5.3^{+3.1}_{-1.4}$ \\
 $ E(\text{Fe\,L}\alpha) [\text{keV}]$ & $0.955^{+0.026}_{-0.044}$ & $0.955^{+0.026}_{-0.045}$ & $0.966^{+0.024}_{-0.042}$ \\
 $ \sigma(\text{Fe\,L}\alpha)[\text{keV}]$ & $0.136^{+0.027}_{-0.019}$ & $0.137^{+0.027}_{-0.019}$ & $0.138^{+0.028}_{-0.019}$ \\
 $ \mathcal{F}_\text{5--60\,keV}^\star$ & $4.495\pm0.010$ & $4.518\pm0.010$ & $4.497\pm0.010$ \\
 $ \text{C}_\text{FPMB}$ & $1.0395\pm0.0023$ & $1.0395\pm0.0023$ & $1.0394\pm0.0023$ \\
 $ \text{C}_\text{XIS\,3}$ & $0.880\pm0.004$ & $0.877\pm0.004$ & $0.885\pm0.004$ \\
 $ \text{C}_\text{PIN}$ & $1.233\pm0.017$ & $1.234\pm0.017$ & $1.230\pm0.017$ \\
 $ \text{C}_\text{GSO}$ & $1.23\pm0.25$ & $1.19^{+0.25}_{-0.24}$ & $1.25^{+0.27}_{-0.26}$ \\
$\chi^2/\text{d.o.f.}$   & 933.14/716& 1006.12/718& 1164.32/719\\$\chi^2_\text{red}$   & 1.303& 1.401& 1.619
\enddata
\tablecomments{$^\dagger$in $10^{-2}$ph\,s$^{-1}$\,cm$^{-2}$~;~$^\ddagger$in $10^{36}$\,erg\,s$^{-1}$ at 10\,kpc~;~$^\star$in keV\,s$^{-1}$\,cm$^{-2}$ } 
\end{deluxetable}

\subsection{Cyclotron line}
\label{susec:crsf}
To describe the CRSF we used a multiplicative line model, with free parameters the centroid energy $E_\text{CRSF}$, width $\sigma_\text{CRSF}$, and optical depth at the line center $\tau_\text{CRSF}$. To investigate the shape of the line in detail, we tried a Gaussian optical depth profile (model \gabs in XSPEC) as well as a pseudo-Lorentzian optical depth profile \citep[model \cyclabs in XSPEC;][]{mihara90a}. Both models have been successfully used to describe the line in \her \citep[e.g., ][]{enoto08a,vasco13a}. \citet{enoto08a} find in \suz data that the Lorentzian line profile describes the data significantly better than the Gaussian profile. For broad lines, both optical depth profiles result in rather similar line shapes.

\begin{figure}
\plotone{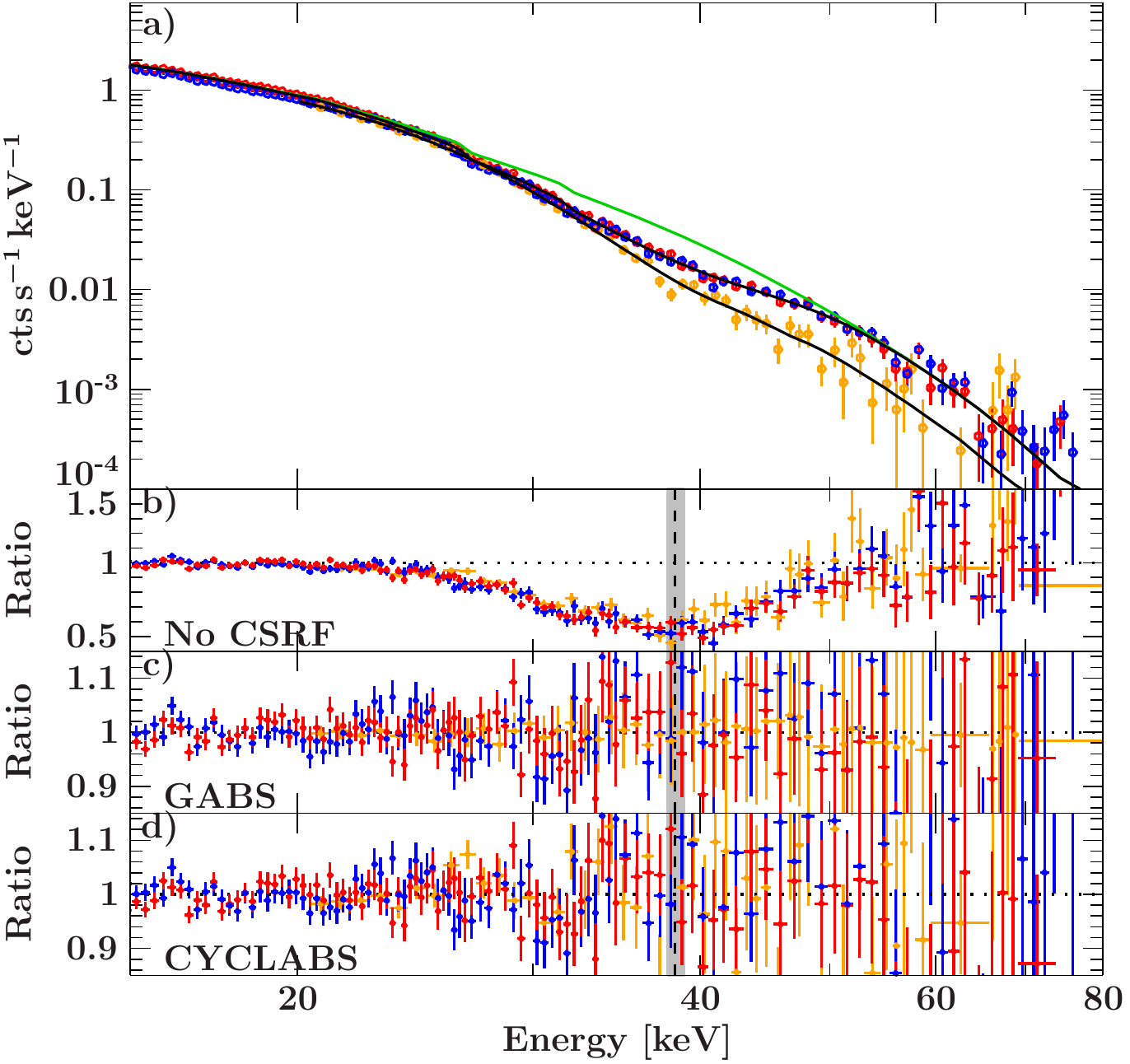}
\caption{\textsl{a)} FPMA (red),  FPMB (blue), and \suz/PIN (orange) spectra of observation I with the best-fit \highe+\gabs model. The green line shows the model without the cyclotron line; \textsl{b)} ratio of the data to the best-fit model with the CRSF turned off; \textsl{c)} ratio to the best-fit model with the \gabs component; \textsl{d)} ratio to the best-fit model with the \cyclabs model. The vertical dashed line indicates the energy of the cyclotron line using the \gabs model, the gray shaded area its 90\% uncertainty interval.}
\label{fig:linshap002}
\end{figure}

\begin{figure}
\plotone{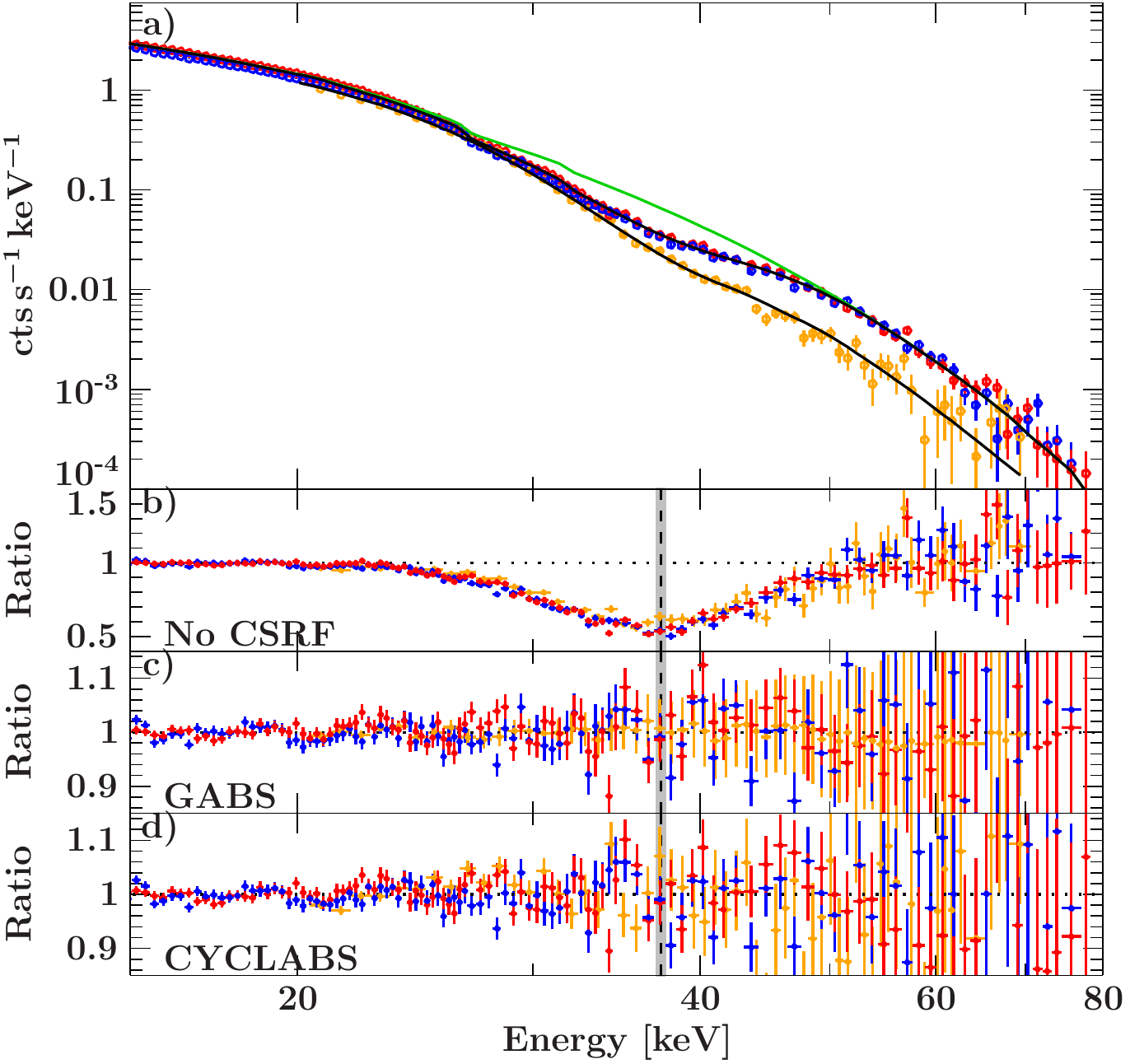}
\caption{Same as Fig.~\ref{fig:linshap002}, but for observation II.}
\label{fig:linshap005}
\end{figure}

\begin{figure}
\plotone{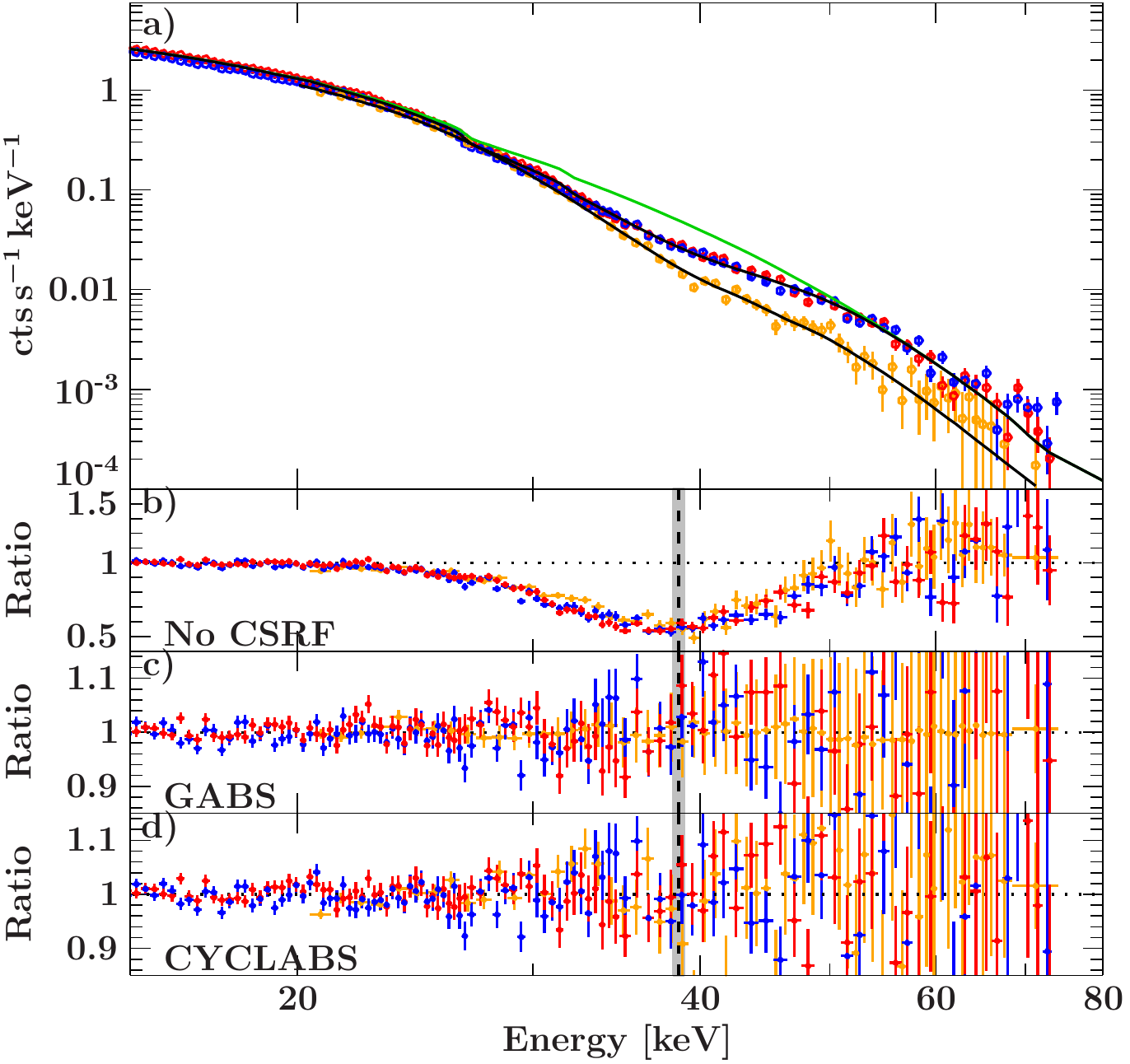}
\caption{Same as Fig.~\ref{fig:linshap002} and Fig.~\ref{fig:linshap005}, but for observation III.}
\label{fig:linshap007}
\end{figure}

We combined both models with the \highe continuum and fitted all three \nustar observations independently. Both models described the data similarly well, with differences in $\chi^2 \leq 5$. We find no evidence for an asymmetric line shape or emission wings at the edges of the CRSF, as shown in Figs.~\ref{fig:linshap002}, \ref{fig:linshap005}, and \ref{fig:linshap007}, for observation I, II, and III, respectively. The choice of the continuum has some influence on the CRSF parameters, but in all cases a Gaussian and a Lorentzian profile fit similarly well. \nustar's FWHM energy resolution at the cyclotron line energy is about 0.5\,keV \citep{harrison13a}. Combined with its very high photon statistics this would allow to measure weak deviations, which would be smeared out by the $\sim$4\,keV energy resolution of PIN.




At the high energy end of the spectrum, we searched for the harmonic CRSF at around twice the energy of the fundamental line, i.e., $\approx$75\,keV. Using observations II and III, as they provides the best \snr, we added a second line with a Gaussian optical depth profile. We fixed the width and energy to twice the corresponding parameter of the fundamental line and only let the depth vary. We did not find an improvement in the fit and obtained an upper limits of $\tau_\text{CRSF,2} \leq 0.16$ for observation I and $\tau_\text{CSRF,2} \leq 0.84$ for observation III. When using an Lorentzian optical depth profile and also fixing width and energy to twice the parameters of the fundamental line, the upper limit in observation III is $\tau^\text{L}_\text{CRSF,2} \leq 0.55$. However, for observation I the fit improved marginally by $\Delta \chi^2 = 9$. The best-fit optical depth is $\tau^\text{L}_\text{CRSF,2}=0.7^{+0.5}_{-0.4}$. This is consistent with  the upper limit of $\tau^\text{L}_\text{CRSF,2}\leq 1.5$  given by \citet{enoto08a}. The improvement in $\chi^2$ is formally significant, but only when fixing the width and energy. Rigorous calculations of the cyclotron scattering, however, show that the harmonic is not necessarily at twice the energy of the fundamental, adding additional systematic uncertainties \citep{harding91a}. Leaving all parameters of the harmonic line free resulted in an unconstrained width of the line. 
In observation I results are similar to observation III, but due to the lower \snr, less constraining. The \nustar data thus put very strong limits on the depth of the harmonic CRSF.

%

\subsection{Soft components}
\label{susec:soft}
\her is known to show a broad iron line, presumably superposing Doppler broadened lines from multiple ionization states. We describe this region by adding two Gaussian lines, Fe\,K$\alpha_\text{n}$ and Fe\,K$\alpha_\text{b}$, a narrow and a broad one, with slightly different energies. A single Gaussian was not sufficient to describe the complex region adequately, mostly constrained by the higher resolution of \suz/XIS. The Gaussians were allowed to vary in energy between 6.4 and 6.9\,keV, to enable modeling of ions up to H-like iron. For observation III we had to freeze the width of the broad component, $\sigma(\text{Fe\,K}\alpha_\text{b})$ to 0.9\,keV, as it was unconstrained by the fit. The combination of the Doppler broadening and different ionization states makes the Fe-line region highly complex.

  A strong line around 1\,keV is evident in the XIS data, which we associated with a  Ni\,IX fluorescence  complex and a possible Fe~L-line at the same energy \citep[see, e.g.,][]{ramsay02a, jimenez-garate02a}. We described this region with an additional broad Gaussian line, as 
 the spectral resolution of XIS is not sufficient to disentangle different narrow lines.

In the XIS data, an additional soft excess is visible, which can be described with a black-body component (\texttt{bbody}) with normalization $A_\text{BB}$ and temperature $kT$, likely originating from the hot accretion disk. The best-fit value corresponds to a radius of $\sim250$\,km for the black-body, similar to the values found by \citet{endo00a}. This feature is consistently seen in all soft X-ray observations at similar temperatures, $kT\approx130$\,eV; see, e.g., \citet{ji09a, oosterbroek01a}. In the \fdcut model the  black-body temperature $kT$ was not constrained by the fit, so we fixed it to 140\,eV, the average value given by the two other models. 

We carefully checked for the presence of absorption  in excess of the expected Galactic absorption of $\approx 1.7\times10^{20}\,\text{cm}^{-2}$ \citep{kalberla05a}. When adding an absorber, the fit only improves marginally and not significantly in terms of $\chi^2$. Using a partially covered absorber resulted in a similar fit, with the covering fraction being consistent with 1.
It is conceivable that outside of the dips in the lightcurve the view on \her is almost unobstructed, so that no additional absorption column was necessary.

\subsection{Results}
\label{susec:phavg_results}

The applied model can be written as
\begin{equation}
\text{ABS} \times \left( \text{CONT} \times \text{CRSF} + \text{BBODY} + \text{fluorescence lines}\right)
\end{equation}
where ABS is the Galactic absorption \citep[using the \texttt{wilm} abundances, ][]{wilms00a} and CONT is either the \highe, \npex, or \fdcut  continuum model. For the remainder of the paper we used only the \gabs model for the CRSF. The best-fit parameters are shown in Tables~\ref{tab:phasavg_002}, \ref{tab:phasavg_005}, and \ref{tab:phasavg_007}, including the flux for the FPMA calibration between 5--60\,keV, $\mathcal{F}_{\text{5--60\,keV}}$. The spectra and residuals of the respective best-fit models for observation II are shown in Fig.~\ref{fig:spec_highe_005}.

The spectral parameters of the \highe model evolve to a slightly softer spectrum with less curvature towards the end of the main-on, with the cutoff energy increasing from  $20.8^{+0.5}_{-0.6}$\,keV to $22.36^{+0.30}_{-0.39}$\,keV  and the folding energy $E_\text{fold}$ going down from  $10.12^{+0.29}_{-0.27}$\,keV to $9.76\pm0.19$\,keV. The photon index $\Gamma$ stays almost constant within the uncertainties. The variations are, however, only marginally significant and should not be interpreted in a physical sense, as the phase-averaged spectrum is a superposition of different spectra from different parts of the accretion column. The slight changes in the spectral parameters reflect the changes seen in the energy resolved pulse-profiles.

 From the phase-averaged fits there is no evidence for a trend in the CRSF line parameters with 35\,d phase, common to all continuum models. Observation II seems to show a slightly lower cyclotron energy and narrower width, but only with marginal significance, and observation I and III have very similar values. As these parameters depend strongly on the local magnetic field, changes should only be investigated in a narrow pulse phase band, see Sect.~\ref{susec:peakspec}.

The broad component of the iron line seems to stay constant in energy and width, but decreases significantly in flux over the three observations. In contrast, the narrower line at higher energies ($\approx 6.6$\,keV) increases in width and flux with 35\,d phase. As the iron line complex is not resolved by either instrument, a physical interpretation is difficult, but as the line at higher energies increases with 35\,d phase, it is indicative of an increased ionization state for the fluorescent medium.

\section{Phase-resolved spectroscopy}
\label{sec:phasresspec}
With the rotation of the neutron star, different regions of the accretion column with different physical conditions will be visible, resulting in strong changes of the spectrum. This effect is observed in all accreting neutron stars, but is especially prominent in \her, as prior studies have shown \citep[e.g.][]{klochkov11a, vasco13a}. 
Most importantly, the energy of the CRSF varies smoothly by a factor of two over the pulse, following the pulse profile \citep{vasco13a}.
Drastic changes in the continuum can happen within less than 0.05 in phase, e.g., the photon index $\Gamma$ drops suddenly around the peak of the pulse profile. Narrow phase-bins with high \snr are therefore required for a physical interpretation of the data.

\begin{figure}
\centering
\includegraphics[width=0.9\columnwidth]{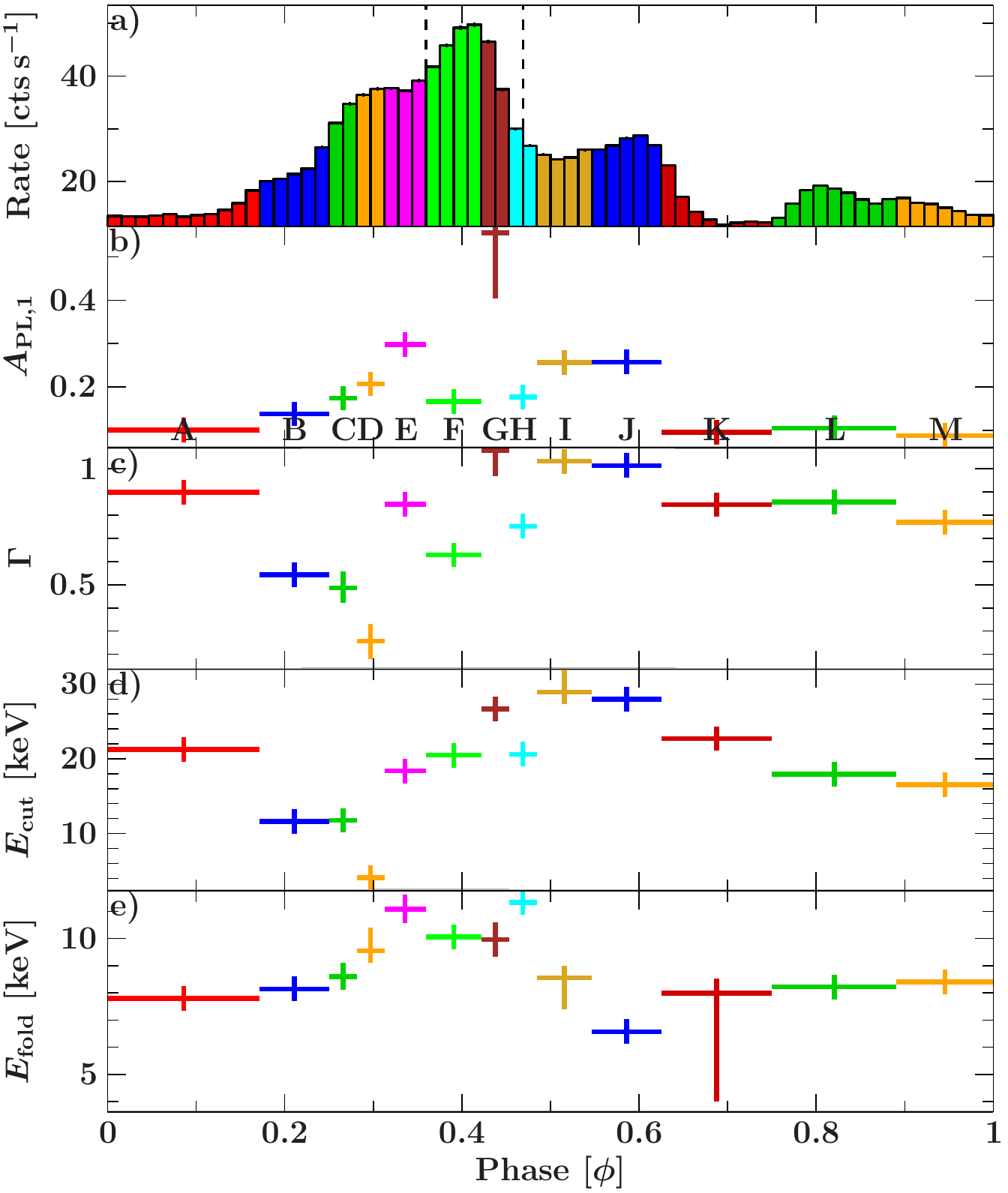}
\caption{Results of the phase-resolved analysis for observation II. \textit{a)} pulse profile between 9-13\,keV, \textit{b)} Normalization of the powerlaw continuum, \textit{c)} photon index, \textit{d)} cutoff energy, \textit{e)} folding energy. The gray line in each panel is the pulse profile in the 9--13\,key energy band. Colors are used to help identifying the phase-bins in all panels and the following plots.}
\label{fig:phasres_cont}
\end{figure}

From the \nustar data extraction we obtained a spectrum in all 64 phase-bins of the pulse profile. To obtain a sufficient \snr, i.e., at least $10^5$\,cts per spectrum, however, we added up to 11 phase-bins for the fit. During sharp changes of the hardness ratio and the pulse profile, only two phase-bins were combined to capture the spectral changes in detail. Overall, \nustar spectra for 13 phase-bins were used, as indicated in Fig.~\ref{fig:phasres_cont}\textsl{a}. As the \suz/PIN showed a lower \snr, we did not use them for phase-resolved spectroscopy.

As the time-resolution of XIS is only 2\,s, no phase-resolved spectroscopy is possible with those data. Therefore, the phase-resolved spectra lack the  soft X-ray coverage of the phase-averaged spectra, and neither the black-body component nor the Fe L-complex around 1\,keV can be investigated. Likewise, constraints on the absorption column and iron line are likely to be less stringent.

To fit the phase-resolved spectra of observation II we used the \highe model, as it describes the phase-averaged spectrum best and is also commonly used in the literature. We only used one Gaussian line to describe the iron line complex, as \nustar's FWHM energy resolution of 0.4\,keV at 6\,keV is not sufficient to resolve both components.  

As presented in Fig.~\ref{fig:phasres_cont}, the shape of the continuum changes dramatically over the pulse phase. Both the photon index $\Gamma$ and the cut-off energy $E_\text{cut}$ vary by more than a factor of two. The photon index drops to its lowest (i.e., hardest) value around the left shoulder of the main peak. There is also a drop in $\Gamma$ shortly after the main peak. 
 The cutoff-energies vary more smoothly over the pulse, but roughly follow $\Gamma$. As high values of $E_\text{cut}$ result in harder spectra, these changes somewhat compensate the hardening of the photon index, and result mainly in modifying the shape and curvature of the spectrum. The folding energy is less variable with pulse phase and typically has higher uncertainties. Overall remarkable periodic spectral changes are evident with pulse phase.

\begin{figure}
\centering
\includegraphics[width=0.9\columnwidth]{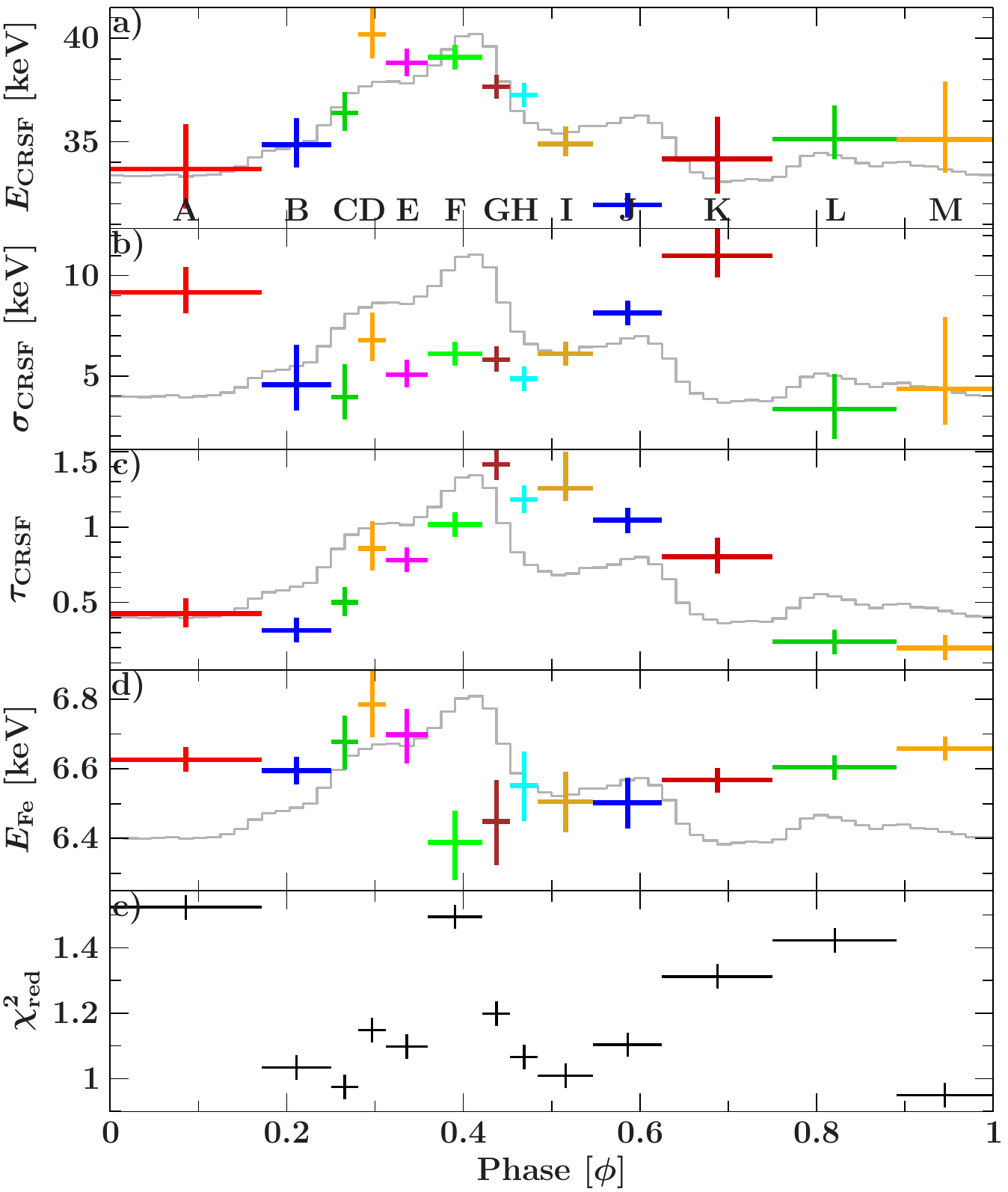}
\caption{Similar to Fig.~\ref{fig:phasres_cont}, but for the parameters of the cyclotron and iron line. \textit{a)} Centroid energy of the cyclotron line, \textit{b)} width of the cyclotron line, \textit{c)} optical depth of the cyclotron line, \textit{d)} energy of the iron line, and  \textit{e)} best-fit \redchi values.}
\label{fig:phasres_crsf}
\end{figure}

For all fits, we froze the width of the \feka line to 0.35\,keV, as it did not change significantly with pulse phase. The centroid energy of the line shows a weak dependence on phase, dropping to its lowest value in phase-bin F,  during the main peak, and rising slowly during the minimum (phase-bins J through M, see Fig.~\ref{fig:phasres_crsf}).

Our data show strong changes of the cyclotron line with phase, 
shown in Fig.~\ref{fig:phasres_crsf}. The line energy, $E_\text{CRSF}$, roughly follows the pulse profile, reaching its highest values around the peak. The optical depth, $\tau_\text{CRSF}$, also is on average higher with higher X-ray flux and peaks during the peak of the pulse profile. However, there seems to be a slight phase shift between the 9--13\,keV flux and the optical depth, with $\tau_\text{CRSF}$ reaching its highest values a little after the peak of the pulse profile.
In all phase-bins, a line with a Gaussian optical depth profile provides a very good description of the line profile. We find no evidence for a significant deviation from a smooth line profile. 

\citet{enoto08a}, using \suz /HXD data, claim a significant detection of the first harmonic CRSF at twice the fundamental line energy, i.e., around 70\,keV, at phases shortly after the main peak. In the \nustar data we find no evidence for that harmonic line. However, this energy is at the very edge of the \nustar sensitive area, so that a weak line might be lost due to weak \snr.

\begin{figure}
\centering
\includegraphics[width=0.9\columnwidth]{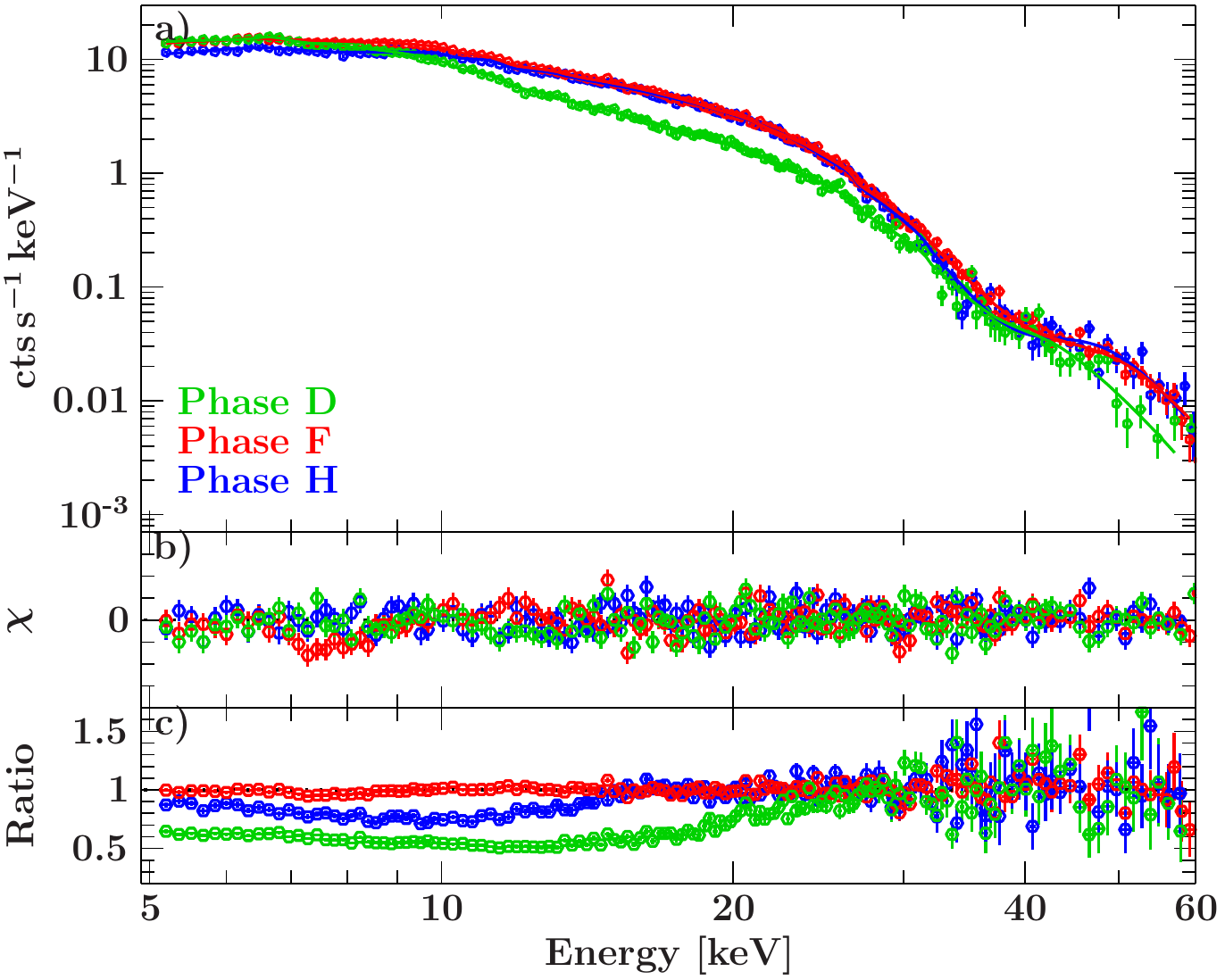}
\caption{\textit{a)} three examples phase-resolved FPMA spectra, of phase-bins C (green), F (red), and G (blue). \textit{b)}  residuals to the respective best-fits. Strong spectral change across the whole energy range are clearly evident. \textit{c)} residuals without including the 10\,keV feature, expressed as a ratio data over model. }
\label{fig:phasres_spectra}

\end{figure}

For phase-bins B, C, D, E, G, H, I, and M it was not possible to obtain a statistically acceptable \redchi value using only the \highe model. Strong residuals remained at energies below 20\,keV. Instead we tried using the \npex model, but did not obtain a better description of the data.
Figure~\ref{fig:phasres_spectra} shows three examples of the phase-resolved spectra where complex changes in the spectral shape below 20\,keV are clearly evident.
 It is possible to obtain an acceptable \redchi value when allowing for a complex absorption, but the column density is highly unconstrained and also often increases to values in excess of $10^{25}$\,cm$^{-2}$, requiring an unphysically high intrinsic luminosity. We thus rule out increased absorption to explain the spectral changes with pulse phase.
   
\begin{figure}
\centering
\includegraphics[width=0.9\columnwidth]{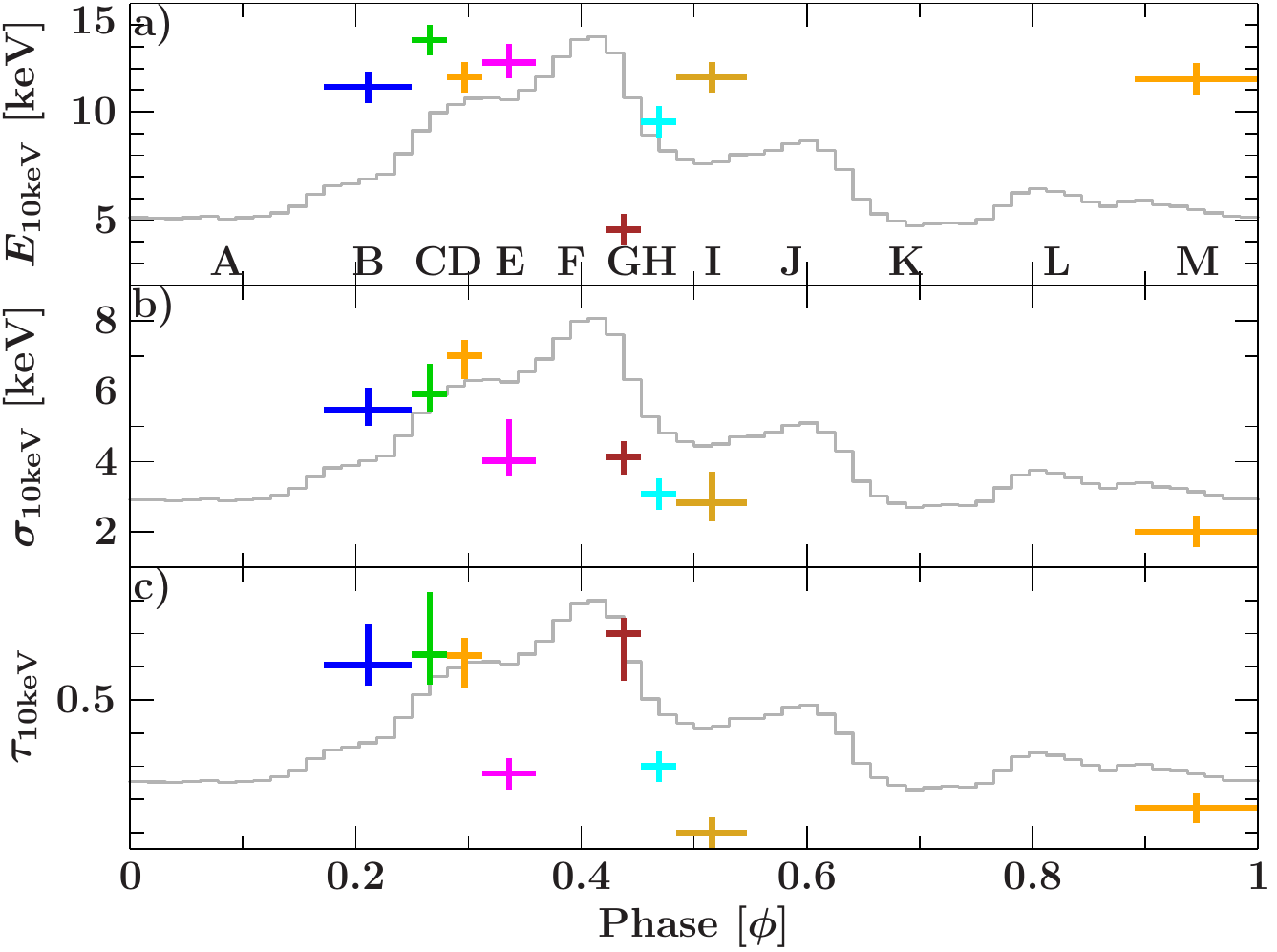}
\caption{Similar to Fig.~\ref{fig:phasres_cont}, but for the parameters of the 10\,keV feature. \textit{a)} centroid energy, \textit{b)} width, and \textit{c)} optical depth.}
\label{fig:phasres_10kev}
\end{figure}

To remove the residuals we instead include another multiplicative line with an energy around 10\,keV and a Gaussian optical depth profile in the \highe model. Although the feature is very strong, it does not significantly influence the measurement of the cyclotron parameters. The width of the feature is at all phases below 7\,keV, so it is completely negligible at the energy of the CRSF (see Fig.~\ref{fig:phasres_10kev}\textit{c}). The \redchi values shown in Fig.~\ref{fig:phasres_crsf}\textsl{d} are for the fits including this feature where necessary. The other phase-bins also have rather unsatisfactory \redchi values, but including an additional feature did not lead to an improvement of the fits or resulted in highly unconstrained spectral parameters. 

This additional feature has a clearly different shape than the calibration feature of the tungsten edge around 10\,keV. The calibration feature shows up as a V-like sharp dip, as seen in the inset of Fig.~\ref{fig:spec_highe_005}. The feature needed for the phase-resolved spectra on the other hand, is very broad, and changes the continuum shape below 20\,keV, as visible in the residuals in Fig.~\ref{fig:phasres_10kev}\textit{c}. We are therefore confident that the small calibration uncertainty does not influence the fit results, and that both features can be clearly separated.

The additional component can be interpreted as a ``10\,keV-feature'', seen in many magnetized and accreting neutron stars \citep[see][and reference therein]{mueller13a}. Even though this feature is seen in many sources, there is no accepted theory for its physical origin thus far.  \citet{vasco13a} describe similar residual structures at the same phases in \xte data of \her. Figure~\ref{fig:phasres_10kev} shows the best-fit values of this feature in the relevant phase-bins.  The ratio between the cyclotron line energy and the energy of the 10\,keV-feature is clearly larger than two in all phase-bins, making it unlikely that this feature is the fundamental line.

\subsection{Pulse peak}
\label{susec:peakspec}
To study the evolution of the spectrum with 35\,d-phase in more detail, we focused on the peak of the pulse profile, removing spectral variance of the pulse profile which could mask variance with 35\,d phase. The phases combined are indicated in Figs.~\ref{fig:pperg_all} and \ref{fig:phasres_cont} and consist of the ones which show the most prominent CRSF. We used the same model as for the phase-resolved spectroscopy, including the 10\,keV feature. Again \suz data were not used. The pulse peak spectra provide high quality data for a specific  viewing angle onto the accretion column.

\begin{deluxetable}{r|lll}
\tablecolumns{4}
\tablecaption{Fit parameters for the peak spectra of all observations.\label{tab:peakres}}
\tablehead{ \colhead{Parameter}  & \colhead{I} & \colhead{II} &  \colhead{III}   }
\startdata 
 $ \Gamma$ & $0.71^{+0.17}_{-0.10}$ & $0.72^{+0.12}_{-0.09}$ & $0.62^{+0.08}_{-0.04}$ \\
 $ E_\text{cut} [\text{keV}]$ & $21.3^{+1.6}_{-9.9}$ & $20.9^{+0.7}_{-0.5}$ & $20.8^{+0.5}_{-0.4}$ \\
 $ E_\text{fold} [\text{keV}]$ & $10.68^{+0.83}_{-0.22}$ & $10.4^{+0.5}_{-0.4}$ & $9.97^{+0.29}_{-0.25}$ \\
 $ E_\text{CRSF} [\text{keV}]$ & $38.6\pm0.6$ & $38.21\pm0.26$ & $38.8\pm0.4$ \\
 $ \sigma_\text{CRSF} [\text{keV}]$ & $6.1\pm0.6$ & $5.53^{+0.35}_{-0.29}$ & $5.9\pm0.4$ \\
 $ \tau_\text{CRSF}$ & $1.13\pm0.10$ & $1.05\pm0.05$ & $1.06\pm0.06$ \\
 $ \sigma_\text{c} [\text{keV}]$ & $2.7^{+1.6}_{-0.7}$ & $2.0\pm0.4$ & $1.76^{+0.27}_{-0.24}$ \\
 $ \tau_\text{c}$ & $0.13^{+0.08}_{-0.05}$ & $0.114\pm0.022$ & $0.130^{+0.017}_{-0.018}$ \\
 $ E_\text{10keV} [\text{keV}]$ & $7.3^{+1.4}_{-1.3}$ & $6.8^{+1.0}_{-0.8}$ & $8.0^{+0.4}_{-0.8}$ \\
 $ \tau_\text{10keV}$ & $0.13^{+0.17}_{-0.07}$ & $0.16^{+0.10}_{-0.08}$ & $0.099^{+0.060}_{-0.021}$ \\
 $ A(\text{Fe\,K}\alpha)^\dagger$ & $0.6\pm0.4$ & $0.80^{+0.32}_{-0.29}$ & $1.0^{+0.6}_{-0.4}$ \\
 $ E(\text{Fe\,K}\alpha) [\text{keV}]$ & $6.38^{+0.12}_{-0.10}$ & $6.46^{+0.07}_{-0.08}$ & $6.56^{+0.10}_{-0.13}$ \\
 $ \sigma(\text{Fe\,K}\alpha)[\text{keV}]$ & $0.21^{+0.15}_{-0.20}$ & $0.32^{+0.10}_{-0.09}$ & $0.36^{+0.17}_{-0.12}$ \\
 $ \mathcal{F}_\text{5--60\,keV}^\star$ & $6.86\pm0.05$ & $9.71\pm0.04$ & $9.90\pm0.05$ \\
 $ CC_\text{FPMB}$ & $0.990\pm0.007$ & $1.044\pm0.004$ & $1.027\pm0.005$ \\
\hline$\chi^2/\text{d.o.f.}$   & 326.94/300& 321.06/308& 317.75/310\\$\chi^2_\text{red}$   & 1.090& 1.042& 1.025
\enddata
\tablecomments{$^\dagger$: in $10^{-2}$ph\,s$^{-1}$\,cm$^{-2}$~$^\star$: in keV\,s$^{-1}$\,cm$^{-2}$ } 
\end{deluxetable}


\begin{table}
\caption{Fit parameters for the peak spectra of all observations, using the spectrum of observation II as basis and modifying it only by Thomson scattering and allowing the \feka-line to vary.}
\label{tab:peakres_cabs}
\centering
\begin{tabular}{r|lll}
   Parameter  & I & II &  III    \\\hline
$N_\text{e} [10^{22}\,\text{cm}^{-2}]$ &   $45.5\pm0.6$ & -- & $\le0.024$  \\
$ A(\text{Fe\,K}\alpha_1)$ & $1.71^{+0.29}_{-0.38}$ &  $0.77\pm0.13$ \\
$ E(\text{Fe\,K}\alpha_1) [\text{keV}]$ & $6.32^{+0.14}_{-0.18}$& $6.46^{+0.07}_{-0.08}$ & $6.50\pm0.09$ \\
 $ \sigma(\text{Fe\,K}\alpha_1)[\text{keV}]$ & $0.69^{+0.31}_{-0.22}$ &  $0.32^{+0.10}_{-0.09}$ & $0.36^{+0.09}_{-0.08}$ \\
\hline$\chi^2/\text{d.o.f.}$   & 370.32/312 & 321.06/308& 528.52/322\\
$\chi^2_\text{red}$   & 1.187 & 1.042 & 1.641\\
\end{tabular}
\end{table}

Table~\ref{tab:peakres} summarizes the results of the spectral fit. We find that the cyclotron line energy is very constant over all three observations. 
The other spectral parameters also do not show a dependence on 35\,d-phase.
We also modeled the spectra using the Lorentzian shaped \texttt{cyclabs} model, as for the phase-averaged spectrum. We did not find a significant difference in the quality of fit. Both cyclotron models describe the line shape very well, without obvious deviations from a smooth line.

\begin{figure}
\centering
\includegraphics[width=0.9\columnwidth]{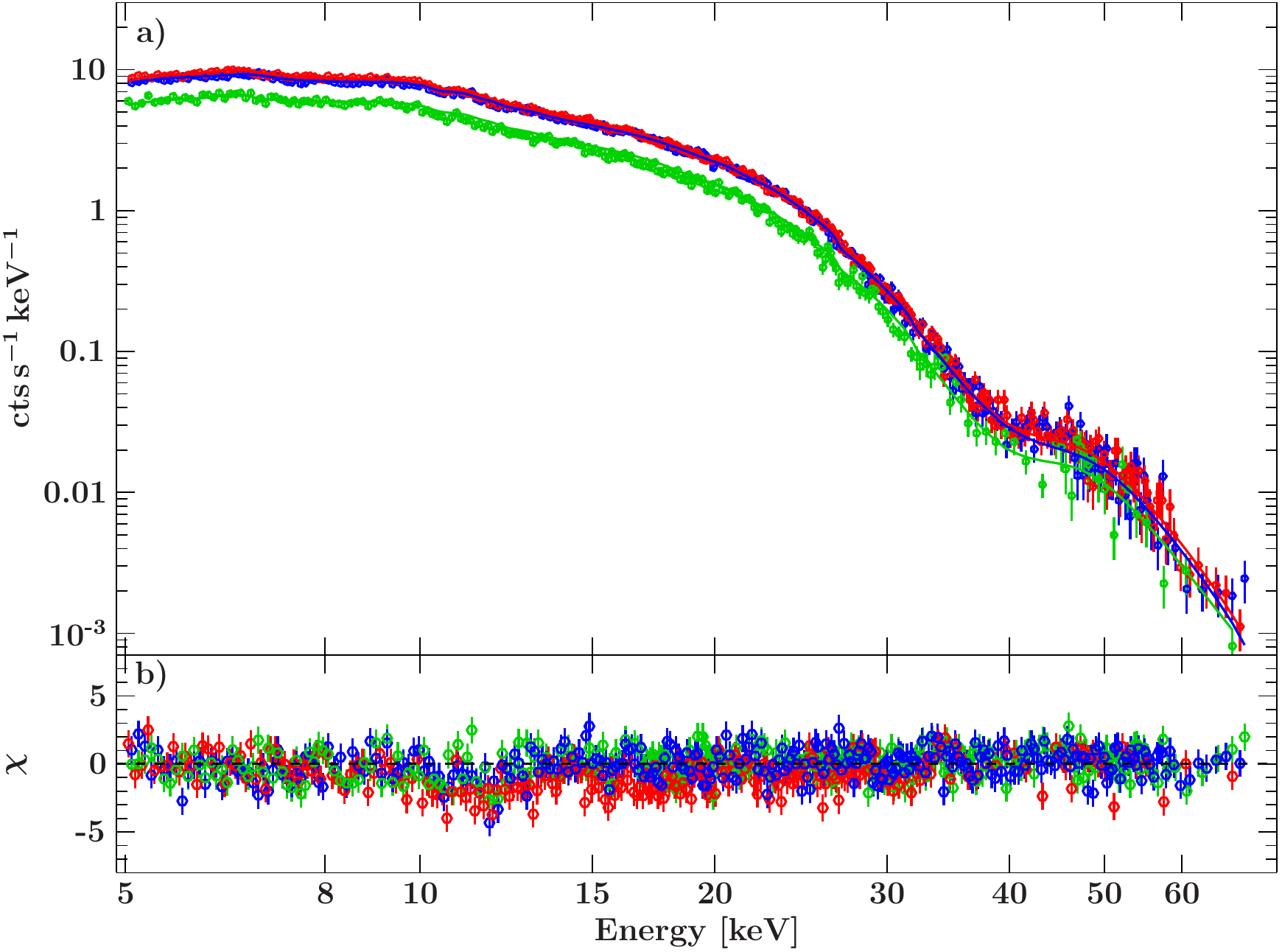}
\caption{\textsl{a)} Pulse peak FPMA spectra of observation I (red), II (blue), and III (green). \textsl{b)} residuals to the respective best-fit model. The residuals around 12\,keV are due to a calibration feature.}
\label{fig:peakspec}
\end{figure}

 Figure~\ref{fig:peakspec} shows all three spectra together, clearly showing that marginal differences exist between them. 
In fact, the changes in spectral shape are so small that all three spectra can be fitted with the same continuum model, simply allowing for additional scattering of photons out of the line of sight. We took the model of observation II as the basic model, as it was in the middle of the main-on. By attenuating this model with Thomson scattering with a column density $N_\text{e}$ (XSPEC model \texttt{cabs})  and allowing for a variable \feka line, an acceptable fit could be achieved for observation I. The best-fit parameters are shown in Tab.~\ref{tab:peakres_cabs}. 
For observation III the fit was clearly worse, as the spectrum shows a slightly different curvature compared to observation II, as indicated by the lower photon index in Tab.~\ref{tab:peakres}. Still, the energy range around the cyclotron line was very well described with the model of observation II.

\section{Summary and Discussion}
\label{sec:outlook}
We have presented the analysis of three data-sets, taken simultaneously with \nustar and \suz, of three different phases of one 35\,d main-on of \her. Our analysis concentrated on the overall spectral shape and the evolution of the CRSF with 35\,d phase and pulse phase. The overall spectral shape is best described with a \highe  model, in which the break at the cutoff energy was smoothed with a multiplicative line. The continuum was modified by a multiplicative absorption line like feature with a Gaussian optical depth to model the CRSF around 37\,keV, as well as fluorescence lines around 6.4\,keV and 1\,keV. 

Using \suz/XIS data, we could extend the energy range down to 0.8\,keV, but we did not find evidence for neutral absorption in excess of Galactic absorption. Previous studies, such as the one by \citet{ji09a} using \chandra                                                                                                                                                                                                                                                                                   data, found very strong absorption columns on the order of $10^{23}$\,cm$^{-2}$, with a partial covering fractions between 0.5-0.9.
Using \xmm data, however,  \citet{ramsay02a} also found only negligible absorption during the main-on, consistent with the \nustar results. These results show that the absorption is variable between different main-ons, and that during the joint \nustar and \suz observation we obtained a relatively clear view of the source. 
In the widely accepted theory that the main-on is caused by the outer rim of the accretion disk moving out of our line of sight, small variations in the thickness of the almost neutral accretion disk and its corona could result in different absorption columns during different main-ons \citep{kuster05a}.

\begin{figure}
\centering
\includegraphics[width=1\columnwidth]{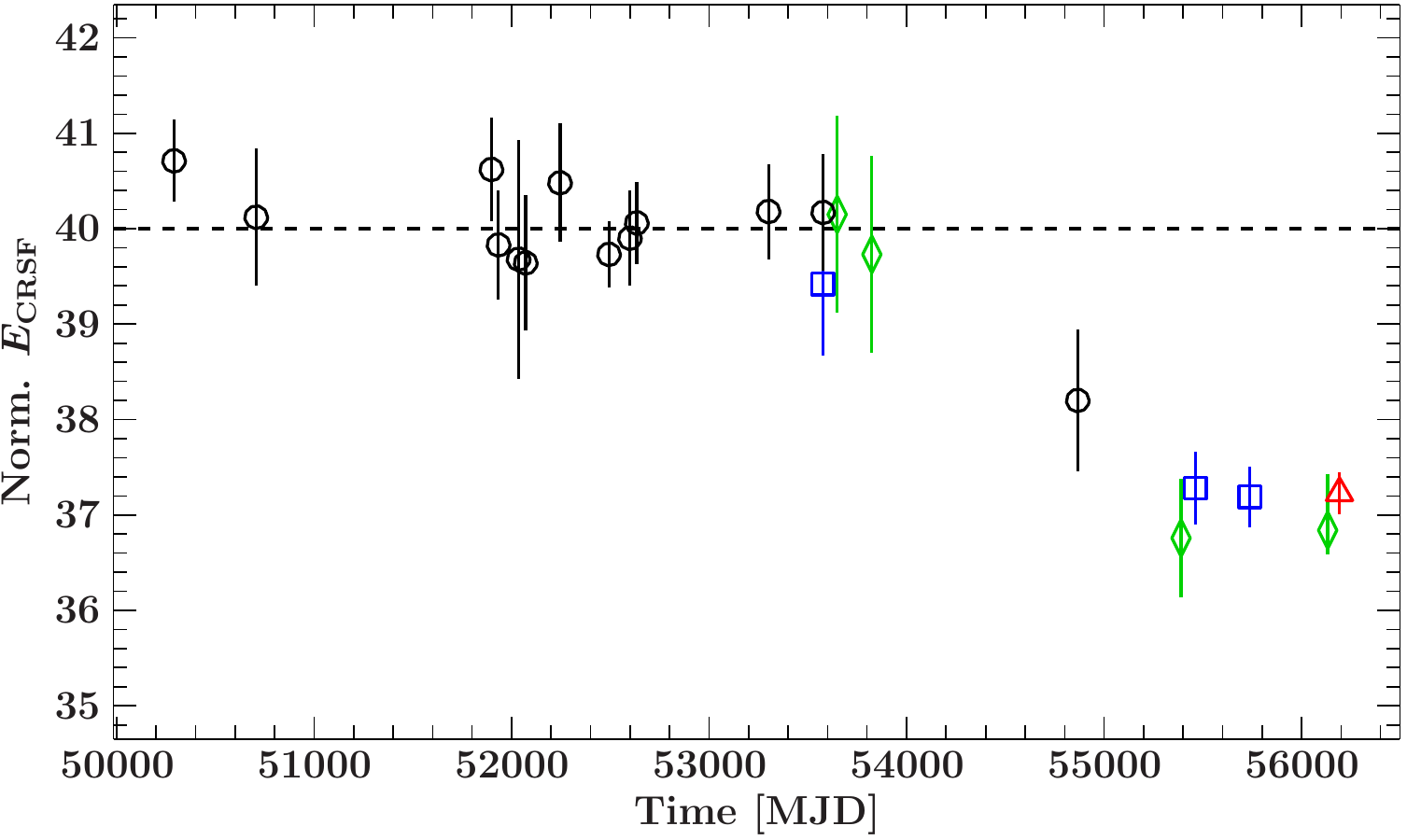}
\caption{Updated version of Fig.~2 (right) of \citet{staubert13b} showing the secular decay of the cyclotron line energy as measured with different instruments over the last 17\,yr, normalized with respect to the average flux. Black circles are data from \xte, blue squares from \inte, green diamonds from \suz, and the red triangle is the \nustar measurement presented here. The \nustar data confirm the decay and provide a highly constrained measurement \citep{staubert13b}.}
\label{fig:crsfdecay}
\end{figure}

The measured values of the cyclotron line energy agree very well with measurements from other missions in the last few years, when normalized to a common luminosity (Fig.~\ref{fig:crsfdecay}). All recent data points are clearly lower compared to measurements taken before 2009 \citep{staubert07a}. To normalize the energy to a common luminosity the known linear correlation between the X-ray luminosity and the cyclotron line energy is used, as described in  \citet{staubert07a}. 
As the reference point, a \xte/ASM countrate of 6.8\,cts\,s$^{-1}$ was chosen. Since the ASM is no longer operational, the monitoring data by \swift/BAT are used for data points after 2011. They are scaled to the ASM rate with the following inter-calibration: ASM [cts\,s$^{-1}$] = 89.0 BAT [cts\,s$^{-1}$]. We checked that ASM and BAT lightcurves agree well enough to be each used as a proxy for the same intrinsic luminosity.
Additionally,  all values in Fig.~\ref{fig:crsfdecay} are taken from 35\,d phases $\varphi_{35}<0.14$, for which the cyclotron line energy dependence on 35\;d phase is minimal \citep{staubert13b}. After taking the luminosity dependence into account, a decline, starting between 2006--2009 becomes visible, with the \nustar measurement providing the latest and best constrained data-point.

 This long-term, luminosity independent decline of the cyclotron line energy cannot be explained by simple models of different altitudes of the line-forming region in the accretion column, as could be caused by changes in the accretion rate, i.e., luminosity. A reconfiguration of the magnetic field, leading to a different geometry of the accretion column, could be responsible for this trend. 
The \nustar data point may indicate that the decay has stopped and the cyclotron energy has settled at a lower lever. The data are, however, also fully consistent with an ongoing decline. For a detailed discussion of this behavior see \citet[in prep.]{staubert13c}.

\subsection{Cyclotron line shape}
The shape of the cyclotron line depends on many factors, including the geometry of the accretion column, the angle between magnetic field and emerging radiation, and the shape of the underlying continuum. As shown by \citet{schoenherr07a}, emission wings around the fundamental line are suppressed when the spectrum is soft, i.e., has a low cutoff energy like seen in \her. We carefully investigated the shape of the cyclotron line in the phase-averaged and pulse-peak spectra and find no evidence for emission wings, or a deviation from a smooth line shape. The lines are described equally well with a multiplicative line with a Gaussian or a Lorentzian optical depth profile. The smooth line shape is in agreement with previous work on \her, where no deviation was found either \citep[see, e.g.][]{enoto08a,vasco13a}. However, \nustar's high energy resolution provides much stronger constraints. 

\citet{staubert07a} and \citet{klochkov11a} found a highly significant positive correlation between the cyclotron line energy and luminosity, clearly showing that the accretion-rate is below the local Eddington limit, where the radiative shock is not strong enough to fully decelerate the in-falling material \citep{becker12a}. Instead, it is stopped by Coulomb interaction further down in the accretion column. The emerging beam pattern in this case is a mix between fan- and pencil-beam.
For this complicated geometry, calculations of the line shape are missing, so that a comparison with theory is not possible at the present time. Calculations for more simple geometries show that a fan-beam pattern produces more symmetric line profiles, with weaker emission wings, compared to a pure pencil-beam profile \citep{schoenherr07a}. 
It seems therefore likely that most of the radiation in \her is emitted through the sides of the accretion column, but updated calculations are needed to assess this interpretation.

\citet{schoenherr07a} also found a strong influence of the accretion column's plasma temperature on line shape. Symmetric lines are only produced, with temperatures below $kT_e \leq 10$\,keV. The thermal distribution of electrons used by \citet{schoenherr07a} is, however, a simplification of the physical conditions in the accretion column. With more detailed simulations becoming available, more reliable constraints will soon be put on the plasma temperature and geometry.

 \subsection{Pulse-phase dependence}
To study the changes of the spectral parameters with the 1.24\,s pulse period, we performed phase-resolved analysis for observation II in 13 phase-bins, dividing the data to capture all spectral changes while at the same time maintaining sufficient \snr. We largely confirm the results found by \citet{vasco13a} using \xte data, and find similarly strong changes of the continuum and cyclotron line parameters. As \xte is no longer operational, \nustar is the instrument best suited for performing such detailed phase-resolved spectroscopy at hard X-rays and measuring the long-term evolution of \her.

 A few phase-bins during the peaks of the pulse profiles required an additional component around 10\,keV to obtain an acceptable fit. We modeled this component with a multiplicative line with a Gaussian optical depth profile. 
Its parameters clearly change with pulse phase, putting its origin somewhere close to the neutron star and its accretion column. As the energy is clearly below half of the cyclotron energy, it cannot be a hitherto undiscovered  fundamental line. 
It is however possible that the 10\,keV feature is not a real physical feature, but an artifact from an imperfect description of the underlying continuum. With the advent of new physical models currently under development \citep[in prep.]{schwarm13a}, this question will be investigated in more detail.

\subsection{35\,d phase dependence}
The spectra of the pulse peak are similar in all three observations and seem to show only a flux dependence on 35\,d-phase. The common model for the 35\,d period in the X-ray flux employs a twisted and tilted accretion disk, which obstructs the X-ray source periodically, leading to a main-on and a short-on \citep{scott00a}. The start of the main-on is defined when the cold, outer rim of the accretion disk moves out of the line of sight and allows a direct view of the X-ray source \citep{leahy02a}. The accretion disk, however, is not thin, but has a definite thickness, with its density following a Gaussian density profile \citep{scott00a}. Furthermore, it is surrounded by a hot corona of ionized gas, scattering X-rays out of the line of sight \citep[and references therein]{shakura99a}.

When the outer rim of the disk moves out of the line of sight, a smooth turn-on is observed. X-rays have to pass through less and less hot and ionized atmosphere as the accretion disk turns. The inferred column density from the \nustar observations at $\varphi_{35}=0.03$ (obs. I) is around $N_\text{e}=4.5\times10^{23}\,\text{cm}^{-2}$  (Tab.~\ref{tab:peakres_cabs}). This fits very well within that picture \citep{schandl94a}. This model is also discussed in detail in \citet{kuster05a} who investigated the turn-on of the main-on and find similar columns, although a little bit earlier in 35\,d-phase than observation I.

We found that the energy of the CRSF does not depend on the 35\,d phase, neither in the phase-averaged spectrum, nor on the pulse-peak data.
In a similar analysis, \citet{vasco13a} found that the peak energy of the line depends on 35\,d-phase, increasing by $\approx 0.7$\,keV per 0.1 in phase. 
They tentatively associated this change with a possible precession of the neutron star, but caution that a quantitative calculation has not yet been done. 
The stability of the \nustar spectral parameters with 35\,d phase (especially the ones of the peak spectra), on the other hand may
indicate that the neutron star does not show free precession. Were it to be precessing,
 a change in the peak spectrum would be expected, similar to the changes seen in phase-resolved spectroscopy.

Free precession of the neutron star was first inferred from the changes of the pulse profile with 35\,d phase \citep{truemper78a}.
In a different model \citep{scott00a}, the changes in the pulse profiles are explained by reprocessing and obscuration of X-rays by the inner region of the accretion disk. However, in this scenario the accretion disk has to come very close to the neutron star. As it is truncated at the magnetospheric radius, this also implies a very small magnetospheric radius of only 20--40 neutron star radii. This is much smaller than the magnetospheric radius inferred from the magnetic field strength as measured by the cyclotron line energy \citep[for a detailed discussion, see][]{scott00a, staubert13a}. The peak of the pulse profile shows the smallest changes with 35\,d phase, in good agreement with the stability of the spectral shape observed by \nustar.

While the \nustar data cannot rule out a freely precessing neutron star,  the observed changes in flux can be explained without it. A varying Thomson scattering column density towards the neutron star is sufficient to explain the small spectral changes. The CRSF is smooth in all observations, limiting possible emission geometries of the accretion column. Rigorous ray-tracing and Monte Carlo simulation in the future will allow us to constrain the geometry of the neutron star and its accretion disk further, especially when comparing to the combined \nustar and \suz data.

\acknowledgments
This work was supported under NASA Contract No. NNG08FD60C, and
made use of data from the {\it NuSTAR} mission, a project led by
the California Institute of Technology, managed by the Jet Propulsion
Laboratory, and funded by the National Aeronautics and Space
Administration. We thank the {\it NuSTAR} Operations, Software and
Calibration teams for support with the execution and analysis of
these observations. This research has made use of the {\it NuSTAR}
Data Analysis Software (NuSTARDAS) jointly developed by the ASI
Science Data Center (ASDC, Italy) and the California Institute of
Technology (USA). We would like to thank John E. Davis for the \texttt{slxfig} module, which was used to produce all figures in this work. FF would also like to thank the Remeis-Observatory Bamberg for their hospitality. JAT acknowledges partial support from NASA Astrophysics Data Analysis Program grant NNX13AE98G. MB was supported by the Centre National d'\'Etudes Spatiales (CNES).

\end{document}